\documentclass[useAMS,usenatbib,usegraphicx]{mn2e}

\usepackage{amssymb}
\usepackage{fixltx2e}

\newcommand{\disp}{\mathbb{D}}
\newcommand{\FAP}{{\rm FAP}}

\title[Keplerian periodogram]{Keplerian periodogram for Doppler exoplanets
detection: optimized computation and analytic significance thresholds}

\author[R.V.~Baluev]{Roman V. Baluev\thanks{E-mail: r.baluev@spbu.ru}\\
Central Astronomical Observatory at Pulkovo of Russian Academy of Sciences, Pulkovskoje
shosse 65, St Petersburg 196140, Russia\\
Sobolev Astronomical Institute, St Petersburg State University, Universitetskij prospekt
28, Petrodvorets, St Petersburg 198504, Russia}

\begin{document}

\date{Accepted 2014 October 19.
      Received 2014 October 17;
      in original form 2014 June 17}

\pagerange{\pageref{firstpage}--\pageref{lastpage}} \pubyear{2014}

\maketitle

\label{firstpage}

\begin{abstract}
We consider the so-called Keplerian periodogram, in which the putative detectable signal is
modelled by a highly non-linear Keplerian radial velocity function, appearing in Doppler
exoplanetary surveys. We demonstrate that for planets on high-eccentricity
orbits the Keplerian periodogram is far more efficient than the classic Lomb-Scargle
periodogram and even the multiharmonic periodograms, in which the periodic signal is
approximated by a truncated Fourier series.

We provide new numerical algorithm for computation of the Keplerian periodogram.
This algorithm adaptively increases the parameteric resolution where necessary, in order to
uniformly cover all local optima of the Keplerian fit. Thanks to this improvement, the
algorithm provides more smooth and reliable results with minimized computing demands.

We also derive a fast analytic approximation to the false alarm probability levels
of the Keplerian periodogram. This approximation has the form $(P z^{3/2} + Q z) W
\exp(-z)$, where $z$ is the observed periodogram maximum, $W$ is proportional to the
settled frequency range, and the coefficients $P$ and $Q$ depend
on the maximum eccentricity to scan.
\end{abstract}


\section{Introduction}

So far, the Doppler radial-velocity (RV) monitoring is one of the most efficient exoplanets
detection methods, both in the number of the planets discovered and in the amount of
information obtained per an individual planet or planetary system.

The first exoplanet discovered by this method, 51~Pegasi~\emph{b}, induced a
single and practically sinusoidal Doppler signal with an amplitude
of approximately $60$~m/s and a period of $4.2$~d \citep{MayorQueloz95}. Thanks to the
continuous growth both of the RV data amount and of the time base, we became able to detect
less massive planets, orbiting their stars at larger distances. Additionally, now we are
frequently dealing with much more complicated planetary systems generating multicomponent,
severely non-linear, and remarkably non-sinusoidal Doppler signals.

Obviously, this progress necessiates the use of considerably more advanced data-analysis
tools then those available in 1995. In this paper we consider the detection of
exoplanets moving along orbits with a large eccentricity. The largest of currently known
exoplanetary orbital eccentricities are above $0.9$, e.g. the well-investigated case
of HD~80606 \citep[e.g.][]{Wittenmyer09}. According to \citet{Schneider} Extrasolar Planets
Encyclopaedia, the largest orbital eccentricity among all exoplanets detected by
radial velocities belongs to HD~20782 with $e=0.97$ \citep{OToole09a}. In Solar System
these eccentricities are typical for comets rather than planets.
From the other side, according there are only three so extreme exoplanets that
reveal $e>0.9$, and the average orbital eccentricity is only about $0.2$ (even after
removal of the hot Jupiters subsample, in which the eccentricities are usually small
or zero). Therefore, high-eccentricity exoplanets are not typical. Nevertheless,
we still have a noticable set of about $50$ exoplanets with rather large eccentricity,
$e>0.6$. This corresponds to $\sim 8$ per cent of the exoplanets that were detected by
radial velocities. This group of exoplanets is the one at which we focus our attention
here.

An exoplanet moving along a highly eccentric orbit induces a drastically
non-sinusoidal Keplerian Doppler signal that cannot be adequately modelled
by a sinusoid. Consequently, the period search methods
like the classic \citet{Lomb76}--\citet{Scargle82} periodogram, as well as any its
extension that still models the putative signal with a plain sinusoid, would likely fail to
detect such a planet, or at least they would not be very efficient. Planets on
highly-eccentric orbits should be more efficiently detected by a periodogram
in which the probe periodic signal is modelled by the Keplerian Doppler function with free
(fittable) orbital parameters. Such a ``Keplerian periodogram'' was originally
introduced by \citet{Cumming04}. Later this periodogram proved rather useful in some
complicated cases involving planets with large orbital
eccentricities \citep[e.g.][]{OToole07,OToole09b}. Thanks to a more accurate model
of the non-sinusoidal planetary RV signal, the Keplerian periodogram allows a more
efficient detection of high-eccentricity exoplanets and more reliable initial determination
of their orbital parameters.

Since typical exoplanetary eccentricities are still not very
large, the Keplerian periodogram should be treated as a specialized tool, rather
than a mass-usage replace for more traditional period-search tools like e.g. the
Lomb--Scargle periodogram. However, this does not remove the need of a special treatment
for high-eccentricity exoplanets on their detection stage. Moreover, the high-eccentricity
exoplanets ($e>0.6$) are more difficult to detect, so their apparently small number can be
due to an observational selection effect in some part. For
example, \nocite{ExoplanetsSeager}\citet{CummingStat} argues that ``there is good agreement
that the detectability falls off for $e>0.5-0.6$''. This only emphasizes the value
of specialized detection tools designed to properly handle large eccentricities.

The Keplerian periodogram did not attain higher popularity due yet another reason. The
Keplerian RV model depends on unknown parameters in a severely non-linear manner. This
forces us to use some iterative non-linear fitting algorithms that increase the computation
complexity dramatically. However, after the computation algorithm is
implemented, the evaluation of an individual Keplerian periodogram is still a feasible task
for modern CPUs. The more difficult issue is that until recently no useful method to
calculate the significance thresholds for such periodograms was available. These thresholds
are necessary to distinguish the real signal from noisy periodogram peaks. Monte Carlo
simulation is no longer an option here, because it needs thousands
of simulated Keplerian periodogram to be processed before we may have a good estimation of
the necessary false alarm probability ($\FAP$). The analytic computation of the periodogram
$\FAP$ is a task that does not have an obvious solution even in the Lomb-Scargle case,
whereas for Keplerian periodograms it is even more difficult.

Although \citet{Cumming04} gave some argumentation concerning the analytic or semi-analytic
approximation of the $\FAP$, after a close investigation, we
find these conclusions unreliable and sometimes even mistaken, because they appear
to implicitly neglect certain importaint non-linearity effects (to be discussed in
more details below). The primary goal of this paper is to construct more
strict approximations to the Keplerian periodogram $\FAP$, involving a
more careful treatment of the Keplerian non-linearity. We achieved this goal by means of
the generalized Rice method, which is an approach of the modern probabilistic theory
of extreme values of random processes and fields. Previously this method
demonstrated a high efficiency in characterizing the significance levels of periodograms
involving linear models \citep{Baluev08a}, and recently it was adapted to periodograms that
involve a general non-linear signal model \citep{Baluev13b}. Basically, now we just need to
substitute the Keplerian model in the general formulae from \citep{Baluev13b},
although this task appeared technically hard.

The advantages of the Rice method are: (i) it is mathematically strict, (ii) it is
very general, (iii) it yields an entirely analytic $\FAP$ estimations, eliminating the need
of any Monte Carlo simulations, (iv) the final $\FAP$ estimations usually can be expressed
by simple elementary formulae, (v) these approximations usually appear
rather accurate, (vi) even if they are not very accurate they still serve as an upper limit
on the $\FAP$, guaranteeing that the actual false positives rate is at least limited by the
desired level. Therefore, this approach remains unbeaten so far, although some promising
fresh results were obtained by fitting the periodogram $\FAP$ with the extreme-value
distributions \citep{Suveges14}.

The structure of the paper is as follows. First of all, in Sect.~\ref{sec_ovw}, we
give a general overview systematizing the family of so-called ``likelihood-ratio
periodograms'', to which the Keplerian periodogram belongs as a special case. In
Sect.~\ref{sec_def} we provide the formal definition of the Keplerian periodogram in a bit
more general formulation than \citet{Cumming04}. In Sect.~\ref{sec_fap} we
describe the application of the Rice method to the Keplerian periodogram and
give the corresponding analytic $\FAP$ estimations. In Sect.~\ref{sec_comp} we describe an
improved computing algorithm for the Keplerian periodogram, involving an optimized sampling
of the space of Keplerian parameters. In Sect.~\ref{sec_demo} we demonstrate this algorithm
using the system of HD~80606 as a rather complicated testcase. In Sect.~\ref{sec_eff} we
provide an analytic comparison between the Keplerian and the sinusoidal models in view
of their signal detection efficiency. In Sect.~\ref{sec_simul}
we perform Monte Carlo simulations to verify the accuracy of the analytic $\FAP$
estimations of Sect.~\ref{sec_fap} and their applicability in practical situations.

\section{Overview of the likelihood-ratio periodograms}
\label{sec_ovw}
In a large part, this work offers a yet another contribution to our series of papers
devoted to the characterization of the periodograms significance levels
\citep{Baluev08a,Baluev09a,Baluev13e,Baluev13b,Baluev13d}. All these papers deal
with the so-called likelihood-ratio periodograms that are based on the
likelihood-ratio statistic comparing two rival models of the data: ``the base model'',
describing the underlying variation expected to be always present in the data, and ``the
alternative model'', expressed as the sum of the base model and of the putative periodic
signal of a given functional form. The detailed mathematical definitions will follow
in Sect.~\ref{sec_def}, and see also \citep{Baluev14c}.

\begin{figure*}
\includegraphics[width=0.99\textwidth]{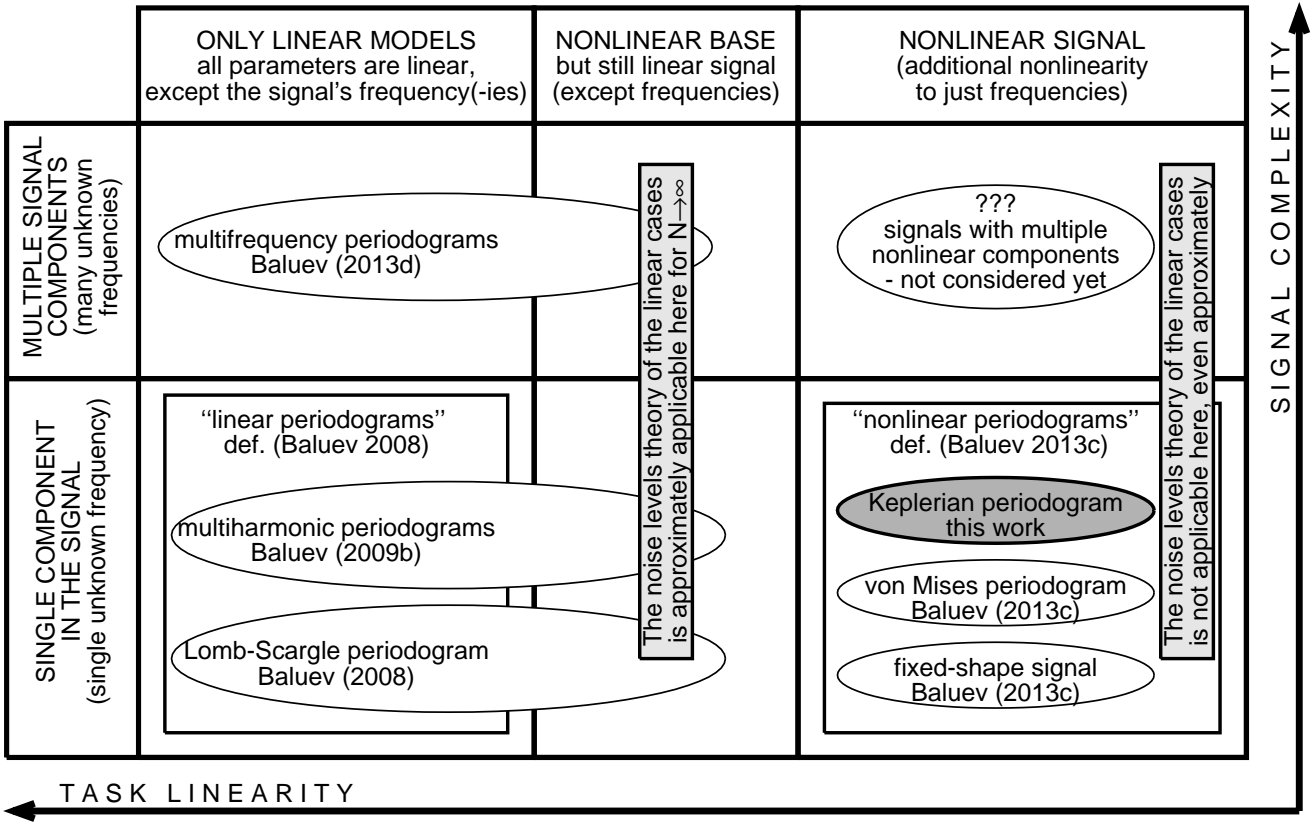}
\caption{Schematic view of the periodograms based on the
likelihood-ratio test, depending on the complexity of the periodic signal to fit and on the
linearity of the fitting task.}
\label{fig_prdg}
\end{figure*}

In view of a large number of periodograms introduced in this work series so far, we provide
a graphical scheme systematizing them based on two properties: the linearity
of the associated maximum-likelihood fitting task and the complexity of the model used
to approximate the probe periodic signal. This is shown in Fig.~\ref{fig_prdg}. We need to
give a few more comments concerning this scheme:
\begin{enumerate}
\item All special linear cases shown in Fig.~\ref{fig_prdg} in the left column
do not involve in their signal model anything more than sinusoids. This model is
either a single sinusoid (e.g. Lomb-Scargle periodogram or its close relative, the
floating-mean periodogram by \citealt{FerrazMello81}) or a sum of sinusoids (the
multiharmonic and multifrequency periodograms). The difference between the multiharmonic
and multifrequency periodograms is how the frequencies of the sinusoids are treated. In the
first case they are binded with each other so that they form the
sequence $f,2f,3f,\ldots,nf$ with only a single basic frequency $f$ to be determined. This
approximates a non-sinusoidal signal by a partial sum of the Fourier series. In the second
case all frequencies are free. In \citep{Baluev08a} a general theory of linear periodograms
was given as well (with only a single fittable frequency). However, so far
we never dealt with a periodic signal modelled by a function more complicated than a sum of
sinusoids, but still linear.

\item There is a qualitative difference between the nonlinear periodograms in which the
nonlinearity is caused by the base model (only) and by the model
of the signal. The nonlinearity in the base model is usually ``weak'' in the sense that the
significance levels can still be approximated by the relevant theory of linear periodograms
(assuming the same signal model and a linearized or even just zero base model). To
symbolically reflect this behaviour, in Fig.~\ref{fig_prdg} we extend the circling
of all the linear cases to the intermediate nonlinearity zone. The nonlinearity in the
signal is essential and a computation of the significance levels for such periodograms
requires a strict treatment of the nonlinear model, e.g. using the methods
from \citep{Baluev13c}. See also a discussion in \citep{Baluev14c}.

\item Another qualitative difference is between a single-component (single unknown
frequency) and multi-component (many unknown frequencies) signals. To strictly
handle the multicomponent variation it is not enough to e.g. extract the signal components
one-by-one, just sequentiall applying a periodogram with the relevant single-component
signal model. To rigorously verify that we did not overestimate the number
of the components in any of possible ways, we should consider the entire ensemble of the
candidate components and separately test the significance of each their subsample.
This requires to apply totally $2^n-1$ periodograms with multicomponent signals,
provided that $n$ is the number of the candidate periodicities. See \citep{Baluev13d} for a
detailed discussion.

\item The most complicated case with nonlinear multicomponent signals is not
investigated yet. What we aim to do in the present work is to consider the ``Keplerian
periodogram'' in which the probe signal is modelled by a \emph{single} Keplerian
RV variation. This still does not provide an entirely rigorous basis for an analysis
of multi-planet systems, because we will assume the classic simplified approach of
extracting the planetary signatures one-by-one. The more rigorous treatment analogous to
the multifrequency analysis \citep{Baluev13d} is a considerably more complicated task in
terms of both the associated theory and numerical computations. We leave this task
of the scope of the paper.

\item All periodograms shown in Fig.~\ref{fig_prdg} belong to the general class of the
so-called ``recursive'' periodograms, as named by \citet{Anglada-Escude12}. In our work we
prefer to call them ``residual'' periodograms, contrary to the traditional periodograms of
residuals. In the cases when the base model is non-trivial and complicated, the residual
periodograms may become considerably more efficient and should be preferred in practice.
See an additional discussion in \citep[Sect.~5]{Baluev14a}.

\item The references shown in Fig.~\ref{fig_prdg} only reflect the works where the
noise levels for a relevant periodogram were well characterized. Some of these periodograms
were actually introduced in earlier papers belonging to other authors. The multiharmonic
periodogram itself was considered by \citet{SchwCzerny96}, while the multifrequency ones by
\citet{Foster95}, and the Keplerian periodogram was introduced by \citet{Cumming04}.
\end{enumerate}

In this paper we do not provide any comparison with the Bayesian model selection
\citep[e.g.][]{Cumming04,Tuomi13}, and in particular with the Bayesian Keplerian
periodogram by \citet{Gregory07a,Gregory07b}. Although some comparison of this
type was initially planned here, the resulting material appeared far more wide
than the reasonable paper size limits would allow, so we plan to release these results in a
separate work in the future.

\section{Keplerian periodograms}
\label{sec_def}
The RV variation induced by an unseen (planetary) satellite moving on a Keplerian orbit is
given by the following formula:
\begin{equation}
\mu = K \{\cos[\omega+\upsilon(\lambda-\omega,e)] + e \cos\omega\},
\label{kepsig}
\end{equation}
with four input parameters: the signal semiamplitude $K$, the mean longitude $\lambda$,
the eccentricity $e$, and the argument of the pericenter $\omega$. The true anomaly
$\upsilon(M,e)$ is a function of the mean anomaly $M=\lambda-\omega$ and of the
eccentricty. Finally, the mean longitude can be represented as $\lambda=\lambda_0 + n
(t-t_0)$, where $\lambda_0$ is the value of $\lambda$ at some reference epoch $t_0$, and
$n$ is the mean motion, which can be tied to the period $P=2\pi/n$ or to the frequency
$f=n/(2\pi)$. Therofore, the signal $\mu$ can be represented as a function of the time $t$
and of five unknown parameters: the signal frequency $f$ and the remaining parameters
$\btheta=\{K,\lambda_0,e,\omega\}$ (with a priori fixed $t_0$). We have separated the
frequency $f$ to comply with the notations of the work \citep{Baluev13b} that we rely upon
below.

In addition to the putative signal~(\ref{kepsig}), the cumulative RV model should also
contain some basic assumptions concerning the task~--- the base or null model. The base
model should at least contain a constant term, and optionally the contributions from some
other (already detected) planets similar to~(\ref{kepsig}). The parameters of planetary
contributions in the base model are approximately known, while the parameters of the
signal are unknown (except for a small number of very wide limits, e.g. the admitted
frequency range). The base model may be even more complicated, e.g. it can be a Newtonian
$N$-body model in some cases. We define the base model as $\mu_{\mathcal
H}(t,\btheta_{\mathcal H})$, where $\btheta_{\mathcal H}$ stores all its parameters
(including e.g. orbital frequencies and other parameters of known planets).

Given the base model $\mu_{\mathcal H}(t,\btheta_{\mathcal H})$, the alternative one
$\mu_{\mathcal K}(t,\btheta_{\mathcal K})=\mu_{\mathcal H}(t,\btheta_{\mathcal
H})+\mu(t,\btheta,f)$ with $\btheta_{\mathcal K}=\{\btheta_{\mathcal H},\btheta,f\}$, and
the input RV time series, we ask: how much the alternative model improves the fit of the
data, in comparison with the base model fit?

Let us define the input time series as $\{t_i,x_i,\sigma_{i,\rm meas}\}_{i=1..N}$ with
$t_i$ being the time of an $i$th observation, $x_i$ being the actual RV measurement, and
$\sigma_{i,\rm meas}$ being its stated (probably incomplete) uncertainty. The measurements
$x_i$ incorporate the random errors $\epsilon_i$, which are assumed independent and
Gaussian. Now we can use several test statistics to compare $\mu_{\mathcal H}$ with
$\mu_{\mathcal K}$. These tests differ by the adopted noise model.

The first group of tests is based on the classic least-square fitting. Let us define the
goodness-of-fit function as
\begin{equation}
\chi^2_{\mathcal H,\mathcal K} = \left\langle (x-\mu_{\mathcal H,\mathcal K})^2 \right\rangle,
\label{chisq}
\end{equation}
where the operation $\langle \phi(t) \rangle$ is the weighted sum of $\phi(t_i)$ taken
with weights $w_i=1/\sigma_{i,\rm meas}^2$ (see \citealt{Baluev08a} for the formal
definition). We may fit the RV models involved by means of minimizing the
function~(\ref{chisq}). In the definitions below we are not interested in the fitted
values of the parameters, but we need the following $\chi^2$ minima:
\begin{equation}
g_{\mathcal H} = \left. \min_{\btheta_{\mathcal H}} \chi^2_{\mathcal H} \right|_{\btheta_{\mathcal H}\approx \btheta_{\mathcal H}^0}, \qquad
g_{\mathcal K} = \left. \min_{\btheta_{\mathcal H}, \btheta, f} \chi^2_{\mathcal K} \right|_{\btheta_{\mathcal H}\approx \btheta_{\mathcal H}^0},
\label{chisqmin}
\end{equation}
where $\btheta_{\mathcal H}^0$ stands for the known initial (approximate) value of
$\btheta_{\mathcal H}$, and the notation $\btheta_{\mathcal H} \approx \btheta_{\mathcal
H}^0$ indicates that the minimizations in~(\ref{chisqmin}) are performed locally over
$\btheta_{\mathcal H}$ (i.e., within the local maximum covering the initial guess). The
minimization over $\btheta$ is global. Based on the minimized $\chi^2$ values, we may
introduce the folowing tests:
\begin{eqnarray}
z = (g_{\mathcal H} - g_{\mathcal K})/2, \qquad
z_3 = \frac{N_{\mathcal K}}{2} \ln \frac{g_{\mathcal H}}{g_{\mathcal K}},
\label{zz3}
\end{eqnarray}
where $N_{\mathcal K}=N-d_{\mathcal K}$ with $d_{\mathcal K} = \dim\btheta_{\mathcal K}$
(similarly we can define $N_{\mathcal H}=N-d_{\mathcal H}$ with $d_{\mathcal
H}=\dim\btheta_{\mathcal H}$). The test statstics~(\ref{zz3}) are direct generalizations
of the linear periodograms $z(f)$ and $z_3(f)$ from \citep{Baluev08a}. The statistic $z$
is designed for the case when $\sigma_{i,\rm meas}$ represent accurate values for the
standard deviations of $\epsilon_i$. This is rarely true. Instead we may use the statistic
$z_3$, which appears when adopting the classic noise model: $\disp
\epsilon_i = \kappa \sigma_{i,\rm meas}^2$, where $\kappa$ is an unknown scale factor
(which is implicitly estimated from the data). This $z_3$ is proportional to the
likelihood ratio statistic associated to the models $\mu_{\mathcal H}$ and $\mu_{\mathcal
K}$. The proportionality factor is $N_{\mathcal K}/N$ and it was introduced mainly to make
the test more conservative in the overfit case, when the dimensionality of the models is
large (in comparison with $N$). The asymptotic behaviour of the statistic $z_3$ for
$N\to\infty$ is the same as for the original likelihood ratio statistic.

However, the classic noise model is inappropriate for exoplanetary Doppler surveys. The
noise variances should be better expressed as $\disp \epsilon_i = \sigma_\star^2 +
\sigma_{i,\rm meas}^2$, where $\sigma_\star$ is the RV jitter \citep{Wright05}. A simple
fitting method handling this noise model, requires some a priori value of $\sigma_\star$,
taken from empiric relations binding $\sigma_\star$ to spectral activity indicators. Given
the value of $\sigma_\star$, we may apply the least square approach above setting $w_i =
1/(\sigma_\star^2+\sigma_{i,\rm meas}^2)$. At present this is probably the most popular
approach, since it introduces only minimum modifications to the least squares method and
thus is easy to implement. However, as we have shown in \citep{Baluev08b}, the estimated
values of $\sigma_\star$ is usually very uncertain and also significantly depends on the
spectrograph and even on the spectrum reduction algorithm used to obtain the actual RV
measurements. In practice it is better to deal with a \emph{fittable} noise model $\disp
\epsilon_i = \sigma_i^2(p) = p + \sigma_{i,\rm meas}^2$, where $p$ is an additional free
parameter. In this case we may use the maximum-likelihood method to estimate the joint
vector of the parameters $\{p, \btheta_{\mathcal H,\mathcal K}\}$. For Gaussian noise,
this likelihood function for the models $\mu_{\mathcal H}$ and $\mu_{\mathcal K}$ should
look like
\begin{equation}
\ln \mathcal L_{\mathcal H,\mathcal K} = - \frac{1}{2} \sum_{i=1}^N \left\{ \ln \sigma_i^2(p)
 + \frac{[x_i-\mu_{\mathcal H,\mathcal K}(t,\btheta_{\mathcal H,\mathcal K})]^2}{\sigma_i^2(p)} \right\} + C,
 \label{lik}
\end{equation}
where $C=N\ln\sqrt{2\pi}$ is a constant. In practice, we prefer to use a modified
likelihood function from \citep{Baluev08b}:
\begin{equation}
\ln \tilde\mathcal L_{\mathcal H,\mathcal K} = - \frac{1}{2} \sum_{i=1}^N \left\{ \ln \sigma_i^2(p)
 + \frac{[x_i-\mu_{\mathcal H,\mathcal K}(t,\btheta_{\mathcal H,\mathcal K})]^2}{\gamma_{\mathcal H,\mathcal K}\,\sigma_i^2(p)} \right\} + C,
 \label{likmod}
\end{equation}
where $\gamma_{\mathcal H,\mathcal K}=1-d_{\mathcal H,\mathcal K}/N$. This modification
involves a preventive bias reduction for the fitted value of $p$: the corrector $\gamma$
basically increases the residuals, which are always systematically smaller than real
errors $\epsilon_i$.

By analogy with~(\ref{chisqmin}) we obtain the relevant likelihood function maxima
\begin{equation}
\tilde l_{\mathcal H} = \left. \ln \max_{p,\btheta_{\mathcal H}} \tilde\mathcal L_{\mathcal H} \right|_{p\approx p^0,\atop \btheta_{\mathcal H}\approx \btheta_{\mathcal H}^0}, \qquad
\tilde l_{\mathcal K} = \left. \ln \max_{p,\btheta_{\mathcal H}, \btheta, f} \tilde\mathcal L_{\mathcal K} \right|_{p\approx p^0,\atop \btheta_{\mathcal H}\approx \btheta_{\mathcal H}^0},
\label{likmax}
\end{equation}
and the associated modified likelihood-ratio statistic from \citep{Baluev08b}:
\begin{equation}
 \tilde Z = \frac{N_{\mathcal K}}{N} \left( \tilde l_{\mathcal K} - \tilde l_{\mathcal H} \right) +
  \frac{N_{\mathcal K}}{2} \ln \frac{N_{\mathcal H}}{N_{\mathcal K}}.
\label{lrZ}
\end{equation}
The offset and normalization of~(\ref{lrZ}) was chosen so that for the classic noise model
$\tilde Z=z_3$, and for $N\to\infty$ the asymptotic behaviour of $\tilde Z$ is the same as
for the conventional likelihood-ratio statistic (i.e. for $Z=\max\ln \mathcal L_{\mathcal
K}-\max\ln\mathcal L_{\mathcal H}$).

Technically, the Keplerian periodogram may be based on any of the three statistics, $z$,
$z_3$, or $\tilde Z$, although $\tilde Z$ is the one preferrable in practice. The
definitions~(\ref{zz3}) and~(\ref{lrZ}) involve the maximizations over all free
parameters, including $\btheta$ and $f$. These statistics represent just some scalar
values that correspond to the maxima of the relevant Keplerian periodograms that we still
need to define. Traditionally, the periodograms are represented as functions of the signal
frequency (or period). Therefore, given some probe frequency, we should compute the
quantities similar to~(\ref{zz3}) and~(\ref{lrZ}), but performing the maximimizations
in~(\ref{chisqmin}) and~(\ref{likmax}) fixing the frequency at the selected value. The
resulting functions of the frequency are our Keplerian periodograms: $z(f)$, $z_3(f)$, and
$\tilde Z(f)$. Below we will rarely use these periodograms themselves, mainly dealing with
their maxima over $f$. Therefore, to prevent further increase in the number of the
notations we will distinguish the Keplerian periodograms (e.g. $z(f)$) from their maxima
(e.g. $z$) only by the dependence on $f$, which in the first case will be always shown
explicitly.

\section{Analytic statistical thresholds for the Keplerian periodogram}
\label{sec_fap}
The parameters $f$, $K$, and $\lambda_0$ of the Keplerian model~(\ref{kepsig}) are present
in the simple sinusoidal model too. Thus, in comparison with the Lomb-Scargle periodogram,
the Keplerian periodogram adds two more degrees of freedom with the parameters $e$ and
$\omega$. Simulatneously, it adds more non-linearity to the task. The only obvious linear
parameter of~(\ref{kepsig}) is the semi-amplitude $K$. There are ways to rewrite this
model so that \emph{two} linear parameters appear \citep{ZechKur09} instead of only a
single $K$. However, the remaining two parameters and the frequency are still non-linear.

It is well known that the statistical significance thresholds for a periodogram are closely
tied to the distributions of the test statistic involved in the periodogram definition. To
compute the $\FAP$ for a detected signal we must assess the distribution of this statistic
under an assumption that the data contain nothing but the underlying variation (given by
the model $\mu_{\mathcal H}$) and noise. \citet{Cumming04} have already considered
this task for the Keplerian periodogram $z(f)$ and some its close relatives. In particular,
he advocated that for $N\to\infty$ the value of the Keplerian periodogram $z(f)$ (with a
fixed $f$) should asymptotically obey the $\chi^2$ distribution with $4$ degrees of freedom
(because we have $4$ free parameters of the model,
except the frequency). But after a preliminary investigation of the task, we find that this
conclusion is likely mistaken.

The asymptotic $\chi^2$ approximation to the distribution of a test statistic like $z$
would appear only if the models $\mu_{\mathcal H}$ and $\mu_{\mathcal K}$ were
both linearizable in the point where we want to compute the distrubution (i.e. at the point
where $\mu\equiv 0$). This is not true for the Keplerian model~(\ref{kepsig}): it cannot be
linearized at $K=0$ without degeneracies. This issue is discussed in more details in
\citep{Baluev13b}, and it originates in the non-identifiability of the Keplerian parameters
at $K=0$. We cannot construct any Taylor decomposition of~(\ref{kepsig}) at $K=0$ that
would be functional for all possible values of other parameters. In other words, the
Keplerian parameters are essentially non-linear here and are similar to the frequency in
this concern. By fixing only the frequency we cannot eleminate or reduce the Keplerian
non-linearity, even in any asymptotic or approximate sense. Therefore, the
$\chi^2$ distribution is not a good approximation here, regardless of whether we
have the frequency fixed or free. Therefore, we do not rely on the results
by \citep{Cumming04} in what concerns the significance estimations.

\citet{Cumming04} have done some Monte Carlo simulations that apparently confirmed his
conclusions about the distributions of the Keplerian periodogram. However, it seems that
his computation of the Keplerian periodogram suffers from undersampling effects that are
discussed below in Sect.~\ref{sec_comp}. This makes his simulation results unreliable. This
is basically the case in which two distorting effects act in opposite directions and
thus largerly compensate and hide each other.

Analytic approximations to the significance levels of the Keplerian periodogram can be
derived using the method introduced in \citep{Baluev13b}. This
method was developed for periodograms based on the chi-square $z$ statistic with
an arbitrary non-linear model of the periodic signal $\mu$, but assuming
that $\mu_{\mathcal H}$ depends on $\btheta_{\mathcal H}$ in a linear manner. This
theory gives the $\FAP$ estimations in the form:
\begin{equation}
\FAP \lesssim M(z),
\label{fapbase}
\end{equation}
with $M(z)$ depending on the structure of the signal model and on the parametric domain.

For example, in the simplest case the signal is given by the sinusoid. Its
harmonic coefficients are unbounded, while the frequency is limited to a segment:
\begin{equation}
0< f < f_{\rm max}.
\end{equation}
This family of ``linear periodograms'', including the classic Lomb-Scargle periodogram, was
considered in \citep{Baluev08a}, where the following was found:
\begin{eqnarray}
\FAP \lesssim M(z) \approx W e^{-z} \sqrt z, \nonumber\\
 W=f_{\rm max} T_{\rm eff}, \quad T_{\rm eff}=\sqrt{4\pi\left(\overline{t^2}-\overline{t}^2\right)}.
\label{lsFAP}
\end{eqnarray}
Here the quantities $\overline{t^k}$ represent the weighted averages
of observation times $t_i$ (taken with the weights appearing in the chi-square function).

In fact, the work \citep{Baluev13b} develops an extension of the same method
to a generalized ``non-linear periodogram'', in which the signal is modelled by an
almost arbitrary non-linear (and non-sinusoidal) periodic function. Now we deal with a
specialized case. The signal is given by the
non-linear model~(\ref{kepsig}), and its parametric domain is defined as:
\begin{eqnarray}
K>0, \quad 0 < \lambda < 2\pi, \quad 0< f < f_{\rm max},\nonumber\\
\quad 0< e< e_{\rm max}, \quad 0< \omega < 2\pi.
\label{dom}
\end{eqnarray}
Here we set upper limits on the frequency and on the eccentricity. The need to set
finite frequency limits is not surprising: similar frequency limits are used for the
Lomb-Scargle periodogram. The eccentricty limit is new. Below it is demonstrated that such
a limit is necessary due to several reasons, including the
singular behaviour of~(\ref{kepsig}) when $e\to 1$. Note that the domain~(\ref{dom})
does not contain pairs of duplicate Keplerian signals, which will be important for the
correctness of the resulting $\FAP$ estimation.

The auxiliary parameter vectors used in \citep{Baluev13b} now look like:
\begin{equation}
\btheta=\{K,\lambda_0,e,\omega\}, \quad \bxi=\{\lambda_0,e,\omega\}, \quad
\bnu=\{e,\omega\}.
\end{equation}
For further convenience we also define $\bxi'=\{f,\lambda_0,e,\omega\}$. Also, we need to
extract the periodic shape function of the Keplerian variation~(\ref{kepsig}), i.e. the
part of $\mu$ that does not depend of $K$:
\begin{eqnarray}
h(t,\bxi') &=& g(\lambda_0 + 2\pi f (t-t_0),\bnu), \nonumber\\
g(\lambda,\bnu) &=& \cos[\omega+\upsilon(\lambda-\omega,e)] + e \cos\omega.
\label{kepsigshape}
\end{eqnarray}
After that we should define a properly normalized model $\psi(t,\bxi')$ such that
\begin{equation}
\langle \psi \mu_{\mathcal H} \rangle\equiv 0, \qquad \langle \psi^2 \rangle \equiv 1.
\end{equation}
The general formula for $\psi$ is given in \citep{Baluev13b}. After that, we need to
calculate the following matrices that describe the local metric of the likelihood function:
\begin{equation}
\mathbfss G_f = \left\langle \frac{\partial \psi}{\partial \bxi} \otimes \frac{\partial \psi}{\partial \bxi} \right\rangle, \qquad
\mathbfss G = \left\langle \frac{\partial \psi}{\partial \bxi'} \otimes \frac{\partial \psi}{\partial \bxi'} \right\rangle,
\end{equation}
corresponding to the cases of fixed $f$ or free $f$, respecitively. Clearly, $\mathbfss
G_f$ is a submatrix of $\mathbfss G$, since $\bxi$ is a subvector of $\bxi'$.

At first, let us apply the approximate approach \citep[sect.~4.1]{Baluev13b}, based on the
assumption of ``unifirm phase coverage''. In this approach the matrix $\mathbfss G$ is
approximated by
\begin{equation}
\mathbfss G \approx \left(\begin{array}{ccc}
4\pi^2 \overline{t^2} q & 2\pi \overline{t} q & 2\pi \overline{t} \bmath v^{\rm T}\\
2\pi \overline{t} q & q & \bmath v^{\rm T} \\
2\pi \overline{t} \bmath v & \bmath v & \mathbfss V \\
\end{array}\right),
\label{matrixG}
\end{equation}
where the quantity $q$, the vector $\bmath v$, and the matrix $\mathbfss V$ are expressed as
\begin{eqnarray}
q = \frac{\overline{{g'_\lambda}^2}}{\overline{g^2}}, \quad
v_i = \frac{\overline{g'_\lambda g'_{\nu_i}}}{\overline{g^2}} -
         \frac{\overline{g g'_\lambda}\,\overline{g g'_{\nu_i}}}{\overline{g^2}^2}, \nonumber\\
V_{ij} = \frac{\overline{g'_{\nu_i} g'_{\nu_j}}}{\overline{g^2}} -
         \frac{\overline{g g'_{\nu_i}}\,\overline{g g'_{\nu_j}}}{\overline{g^2}^2}, \quad
R_{ij} = V_{ij} - \frac{v_i v_j}{q}.
\label{vmv}
\end{eqnarray}
In the last formulae, the function $g$ should be substituted from~(\ref{kepsigshape}), the
over-lines denote the integral averaging over the periodic argument $\lambda$. The function
$g$ should necessarily satisfy here the prerequisite condition $\overline g = 0$, which for
our Keplerian model is already fulfilled. Note that the continuous averaging operation used
in~(\ref{vmv}) is thus different from the discrete averaging
$\overline{t^k}$ from~(\ref{lsFAP}) and~(\ref{matrixG}).

The most hard part of the work is the computation of~(\ref{vmv}). The second term in the
expression for $v_i$ vanishes, because the integral of $(g g'_\lambda)$ over a single
period of $\lambda$ is obviously zero (this useful property was actually missed in
\citealt{Baluev13b}, as it is valid for an arbitrary smooth and $\lambda$-periodic $g$).
But the other terms appearing in~(\ref{vmv}) are non-trivial. The schematic plan of the
computation contains two steps:
\begin{enumerate}
\item Derive the necessary derivatives of the Keplerian shape function $g$. In fact, these
derivatives are already available in various literature \citep{Pal10,WrightHoward09}.

\item Compute the necessary averages of the combinations appearing in~(\ref{vmv}) over
$\lambda$. The derivatives of $g$ involve the functions of the type $r^n \cos k\upsilon$
and $r^n \sin k\upsilon$, where $r$ is the Keplerian radius-vector. Averages of such
expressions can be found in handbooks on the Keplerian motion
\citep[e.g.][]{Kholsh-twobody}.
\end{enumerate}

Regardless of principal feasibility, the manual computation of~(\ref{vmv}) is an extremely
difficult task. We have undertaken a few such attempts, and none of them was successful due
to mistakes appearing in the process. We eventually decided to use the MAPLE
computer algebra system to obtain more reliable expressions for~(\ref{vmv}). The details of
these computations are given in the MAPLE worksheet attached as the online supplement to
the article. After some polishing of the MAPLE results, we obtain the following:
\begin{eqnarray}
q = \frac{(1+\beta^2)^3}{(1-\beta^2)^5} \frac{B(\beta,\omega)}{A(\beta,\omega)}, \nonumber\\
v_\omega = \left(\frac{1+\beta^2}{1-\beta^2}\right)^2 \frac{1}{A(\beta,\omega)}, \quad
v_\beta = \frac{\beta (1+\beta^2)}{(1-\beta^2)^3} \frac{\sin 2\omega}{A(\beta,\omega)}, \nonumber\\
V_{\omega\omega} = \frac{1-\beta^4}{A^2(\beta,\omega)}, \qquad
V_{\omega\beta} = \frac{\beta\sin 2\omega}{A^2(\beta,\omega)}, \nonumber\\
V_{\beta\beta} = \frac{C(\beta,\omega)}{(1-\beta^4) A^2(\beta,\omega)}, \nonumber\\
R_{\omega\omega} = \frac{\beta^2(1-\beta^4)(4+\beta^2)}{A^2(\beta,\omega) B(\beta,\omega)}, \,
R_{\omega\beta} = \frac{\beta^3 (4+\beta^2) \sin 2\omega}{A^2(\beta,\omega) B(\beta,\omega)}, \nonumber\\
R_{\beta\beta} = \frac{(4+\beta^2)(A(\beta,\omega) B(\beta,\omega)+\beta^4\sin^2 2\omega)}{(1-\beta^4) A^2(\beta,\omega) B(\beta,\omega)}, \nonumber\\
\det\mathbfss R = \frac{\beta^2 (4+\beta^2)^2}{A^3(\beta,\omega) B(\beta,\omega)},
\label{vmvkep}
\end{eqnarray}
where
\begin{eqnarray}
A = 1 - \beta^2\cos 2\omega, \quad B = A + \beta^2 (4 + \beta^2), \nonumber\\
C = 4 + 5\beta^2 + \beta^4 - \beta^2 (2+\beta^2) \cos^2 \omega - 4\beta^4 \cos^4 \omega, \nonumber\\
\beta=\frac{e}{1+\eta}, \quad \eta=\sqrt{1-e^2}.
\end{eqnarray}
Here we have replaced the eccentricty $e$ by the new parameter
$\beta$, because these formulae appear more seizable in terms of $\beta$. This might
be suspected e.g. from the formulae of the integrals of $r^n \cos k\upsilon$ in
\citep{Kholsh-twobody}.

Substituting~(\ref{vmvkep}) to~(\ref{matrixG}) and then to the suitable expressions for
$M(z)$ from \citep{Baluev13b}, we obtain the $\FAP$ approximations of the following form:
\begin{eqnarray}
\FAP(z) \lesssim M(z) = \exp(-z) \big[2 z X_f(e_{\rm max}) + \nonumber\\
 + Y_f(e_{\rm max}) \sqrt{\pi z} + \mathcal O(z^0) \big] \quad ({\rm fixed}\ f), \nonumber\\
\FAP(z) \lesssim M(z) = W \exp(-z) \sqrt z \big[2z X(e_{\rm max}) + \nonumber\\
 + Y(e_{\rm max}) \sqrt{\pi z} + \mathcal O(z^0) \big] \quad ({\rm free}\ f).
\label{kfap}
\end{eqnarray}
The functions $X$ and $Y$ are expressed as
\begin{eqnarray}
X_f(e_{\rm max}) &=& \int\limits_0^{\beta(e_{\rm max})} d\beta \int\limits_0^{2\pi} \sqrt{q \det\mathbfss R}\, \frac{d\omega}{2\pi}, \nonumber\\
X(e_{\rm max}) &=& \int\limits_0^{\beta(e_{\rm max})} d\beta \int\limits_0^{2\pi} q \sqrt{\det\mathbfss R}\, \frac{d\omega}{2\pi}, \nonumber\\
Y_f(e_{\rm max}) &=& \left. \int\limits_0^{2\pi} \sqrt{q R_{\omega\omega}}\, \frac{d\omega}{2\pi} \right|_{\beta=\beta(e_{\rm max})}, \nonumber\\
Y(e_{\rm max}) &=& \left. \int\limits_0^{2\pi} q \sqrt{R_{\omega\omega}}\, \frac{d\omega}{2\pi} \right|_{\beta=\beta(e_{\rm max})}.
\label{XY}
\end{eqnarray}
In~(\ref{kfap}), the terms containing $X$ reflect the expected number of local maxima of
the likelihood function in the entire domain~(\ref{dom}), while the terms with $Y$ reflect
the number of local maxima on the boundary of~(\ref{dom}) at $e=e_{\max}$. The terms
corresponding to the boundary $f=f_{\rm max}$ are negligible, because they do not contain
the large factor $W$.

The $\FAP$ approximations~(\ref{kfap}) look rather similar to those for the von Mises
periodogram discussed in \citep{Baluev13b}, with the eccentricity being an analogue of the
localization parameter. And similarly to the von Mises periodogram, it appears that the
coefficients $X$ and $Y$ tend to infinity if $e$ is unbounded, so we must limit the the
eccentricity by some $e_{\rm max}<1$ to have meaningful results. This $e_{\rm max}$ should
be selected a priori, like $f_{\rm max}$.

The integrals~(\ref{XY}) are not elementary, except for $X_f$, for which we obtain
(again with MAPLE):
\begin{equation}
X_f(e_{\rm max}) = \left. \frac{\beta^2 (24-21\beta^2+7\beta^4)}{12 (1-\beta^2)^3} \right|_{\beta=\beta(e_{\rm max})}.
\end{equation}
Most of the other integrals can be represented through pretty unpleasant combinations of
elliptic integrals. Moreover, for $X(e)$ MAPLE obtains an indefinite result due to
some tricky degeneracy appearing in the process. Perhaps this integral may involve
something more complicated than even the elliptic functions. The details are given
in the attached MAPLE worksheet.

In any case, these accurate expressions are difficult for practical use, and
we therefore tried to fit~(\ref{XY}) numerically using some more
simple formulae. After a few experiments, the following semi-empiric expressions
were constructed:
\begin{eqnarray}
X_f(e) &\simeq& 0.5 \varepsilon^2 + 0.1042 \varepsilon^3 - 0.0914 \varepsilon^{2.44}, \nonumber\\
X(e) &\simeq& 0.5 \varepsilon^2 + 0.0350 \varepsilon^6 + 0.3334 \varepsilon^{3.86} + 0.0774 \varepsilon^{5.03}, \nonumber\\
Y_f(e) &\simeq& \varepsilon + 0.5033 \varepsilon^3 + 0.2585 \varepsilon^{2.44}, \nonumber\\
Y(e) &\simeq& \varepsilon + 0.3125 \varepsilon^6 + 2.3725\varepsilon^{3.05} + 0.9868\varepsilon^{4.86},\nonumber\\
& & \varepsilon = \frac{e}{\eta} = \frac{2\beta}{1-\beta^2}.
\label{XYapprox}
\end{eqnarray}
These approximations preserve the asymptotic behaviour of~(\ref{XY}) for $e\to
0$ and for $e\to 1$, and their relative errors are below $1$ per
cent for $X$ and $Y$ and below $5$ per cent for $X_f$ and $Y_f$. It follows that for $e\to
1$ the $\FAP$ increases as either $\eta^{-6}$ (for the free $f$) or $\eta^{-3}$ (for the
fixed $f$).

As we can see, although the procedure of computation was very hard in its internals, the
final approximations~(\ref{kfap}) and~(\ref{XYapprox}) are not that complicated, and
even became elementary. Also, one may note that the $\FAP$ of the fixed-frequency case
clearly does not match the $\chi^2$ distribution with $4$ degrees of freedom. This $\chi^2$
distribution would imply $\FAP(z) = (z+1) \exp(-z)$, and this is more or less close to the
first formula of~(\ref{kfap}) only when $X_f=0.5$, achieved for $e_{\rm max}\approx 0.7$.

The comparison of these theoretic results with the results of Monte Carlo simulations will
be given in Sect.~\ref{sec_simul} below.

\section{Keplerian periodogram computation}
\label{sec_comp}
Clearly, fitting the data with the non-linear model~(\ref{kepsig}) is more complicated
than fitting e.g. the sinusoidal model. It follows from \citep{Cumming04,ZechKur09} that
the likelihood function often has multiple maxima in the Keplerian case, and these maxima
concentrate at high eccentricities. When computing the Keplerian periodogram, it is
important to process \emph{each} of these local maxima. Missing even a single such local
maxima may result in underestimated likelihood-ratio statistic, and hence in a decreased
detection power. This may also generate discontinuties in the graph of the Keplerian
periodogram, occuring when the global likelihood maximum moves to a missed local one. Such
discontinuties can be noticed in the plots by \citet{ZechKur09}, and this indicates that
the Keplerian parametric space might be undersampled in that work. It seems that
\citet{ZechKur09} used some regular and uniform grid of the parameters for maximization,
although they recognize that a non-uniform grid with the density increasing with $e$ would
be preferred. \citet{Cumming04} were ``trying several initial starting values for the
phases and eccentricity''. This likely means that the parametric space was undersampled
too, as only ``several'' initial conditions is often too small in this task.

The algorithm that we propose here tries to achieve the best performance without
sacrificing the safety of the final result (that is, disallowing to loose local maxima).
It is based on the following principles:
\begin{enumerate}
\item Like \citet{Cumming04}, we use the non-linear Levenberg-Marquardt optimization,
subsequently trying starting conditions from a relatively rarified set (roughly one or a
few points per each local maximum of the likelihood function). The plain grid-scanning
approach used by \citet{ZechKur09} would require a much more dense grid, since they needed
to also sample each local maximum at a high enough density.
\item In order to not miss any local maximum and avoid undersampling, we pay more
attention to the construction of the grid of starting conditions. This grid is not
uniform: its density adaptively increases together with the expected density of the local
maxima.
\item Our algorithm does not currently use the main idea of the method by
\citet{ZechKur09} to extract two linear parameters in the model~(\ref{kepsig}) and treat
them separately from the remaining non-linear ones. Neither we use the similar approach
proposed by \citet{WrightHoward09} for RV curves fitting.
\end{enumerate}

Now our task is to construct an optimal multi-dimensional grid that would be rarified as
much as possible, still disallowing any likelihood maxima to evade between the grid nodes.
From \citep{ZechKur09} and from Sect.~\ref{sec_fap} above we may conclude that local maxima
of the likelihood function concentrate at high eccentricities, and this is confirmed by the
formulae~(\ref{kfap},\ref{XYapprox}). The reason for such behaviour is
that for large $e$ the Keplerian signal~(\ref{kepsig}) is almost constant most of the time,
except for a short periastron passage events. The characteristic time spend by the
planet near its orbital pericenter is inversely proportional
to the pericentric angular velocity of the planet. From the second
Kepler's law the planetary angular velocity can be determined as $\dot\upsilon = f (a/r)^2
\eta$. This turns into $f \eta/(1-e)^2$ in the pericenter. Therefore, the planet
spends near the pericenter roughly $(1-e)^2/\eta = \eta^3/(1+e)^2$ fraction of
each its orbital period. This time decreases for large $e$ as $\sim \eta^3$. To adequately
trace such short spikes in the Keplerian radial velocity, some of the Keplerian parameters
have to be sampled at an increasingly high density, when $e$ tends to unity.

The optimal grid can be constructed using the results of Sect.~\ref{sec_fap}. The integral
for $X$ in~(\ref{XY}) is proportional to the expected number of the local maxima
of the likelihood found inside the integration domain, while the integrand
is proporional to the local density of these maxima. Therefore, the ideally optimal grid of
Keplerian parameters should comply with the following probability density function (PDF):
\begin{equation}
p_{f\lambda\beta\omega} \propto \sqrt{\det\mathbfss G} = q\sqrt{\det\mathbfss R} =
\beta (4+\beta^2) \frac{(1+\beta^2)^3}{(1-\beta^2)^5} \sqrt{\frac{B(\beta,\omega)}{A^5(\beta,\omega)}}
\label{peakdens}
\end{equation}
It is clear that:
\begin{eqnarray}
p_{f\lambda e\omega} = p_f p_\lambda p_{e\omega}, \qquad p_f = \frac{1}{f_{\rm max}}, \quad p_\lambda = \frac{1}{2\pi}, \nonumber\\
p_{e\omega} = \frac{q \sqrt{\det\mathbfss R}}{2\pi X(e_{\rm max})} \frac{d\beta}{de}, \quad e\leq e_{\rm max},
\end{eqnarray}
implying that $f$ and $\lambda_0$ are uniformly-distributed and independent from each other
and from $e$ and $\omega$, while $e$ and $\omega$ are mutually correlated. It follows that
the marginal PDF and cumulative distribution function (CDF) of the eccentricity in the grid
are given by
\begin{equation}
p_e = \frac{X'(e)}{X(e_{\rm max})}, \quad P_e = \frac{X(e)}{X(e_{\rm max})}, \quad e\leq e_{\rm max},
\end{equation}
and the biparametric PDF of $e$ and $\omega$ can be expressed as
\begin{equation}
p_{e\omega} = p_e p_{\omega|e}, \qquad p_{\omega|e} \propto \sqrt{\frac{B(\beta(e),\omega)}{A^5(\beta(e),\omega)}}.
\end{equation}

We must emphasize that this approach necessarily requires a \emph{random} rather
than a regular grid. Even if $\lambda_0$ and $f$ are independent from other parameters, we
cannot just select two indepenent grids for $\lambda_0$ and $f$, some grid for $e$ and
$\omega$, and form the resulting multidimensional grid as a Cartesian product of these
three. In such a case the values of $\lambda_0$ and $f$ would attain the same values for
all $e$ and $\omega$, but this may lead to lost local maxima. To achieve
an optimal grid, the values of $\lambda_0$ and $f$ should be sampled anew for each new pair
$(e,\omega)$. This in fact implicitly simulates an increasingly more dense distribution of
the sampled values of $\lambda_0$ and $f$ when $e$ grows: in a vicinity of a larger $e$ the
number of grid nodes is considerably larger, implying a smaller $\lambda_0$- and
$f-$separation between neighbouring nodes. In fact, some increasing of the phase and
frequency resolution for larger $e$ is a mandatory property of the required grid.

However, a disadvantage of a random grid is that
it makes the computation results random. There is no guarantee that random nodes
do not accidentally avoid some regions of the parametric space, potentially leading to lost
local maxima. To suppress this effect we may increase the number of the generated nodes,
but this makes the grid oversampled in other
regions, slowing the periodogram computation down.

We therefore still need to have a regular grid, but this requires to correctly simulate the
increase of the density in $\lambda_0$ and $f$. Note that for $e\to 1$ we have $p_e \sim
\eta^{-8}$, which is an excessively quick growth. The local density of the
likelihood function peaks along an $e$-isoline (fixing all parameters but $e$) can
be computed by restricting the matrix $\mathbfss G$ to its single element $V_{\beta\beta}$,
given in~(\ref{vmvkep}). The necessary density (for $\beta$) is proportional
to $\sqrt{R_{\beta\beta}}$. Mapping it from $\beta$ to $e$, we may obtain that the
reasonable grid density along $e$-isolines should scale as $\sim
\eta^{-5/2}$ (the pessimistic case with $\omega$ near $0$ or $\pi$) or
as $\sim \eta^{-3/2}$ (the optimistic case with $\omega$ near $\pm \pi/2$). We adopt the
average rate of $\nu_e(e)\sim \eta^{-2}$. Since the original growth rate was $\sim \eta^8$,
for each grid layer with fixed $e$ we still have about $\sim \eta^6$ values of $\lambda_0$
and $f$ to distribute uniformly and independently. The reasonable local density of the grid
nodes along an $\omega$-isoline is given by $\sqrt{R_{\omega\omega}}$, and their total
number is proportional to $\int\limits_0^{2\pi} \sqrt{R_{\omega\omega}} d\omega = 2\pi$.
Expectedly, the total number of the grid nodes along an $\omega$-isoline should remain
constant for all $e$.

The most obvious way is to generate $\sim \eta^{-3}$ grid nodes per each
$\lambda_0$- and $f$-isoline (for a given $e$). This agrees with the expected peaks number
along the isolines obtained after restriction of $\mathbfss G$
to the corresponding diagonal element and integration of its square root over the selected
parameter. Both for the $\lambda_0$- and $f$-isolines, we have the number of peaks
proportional to $\sqrt q$, corresponding to the growth rate of $\sim
\eta^{-5/2}$ (optimistic cases $\omega=0$ or $\pi$) to $\sim \eta^{-3}$ (pessimistic cases
$\omega=\pm \pi/2$). Our grid corresponds to the pessimistic case here. The total number of
the grid nodes behaves as $\int\limits_0^{e_{\rm max}} \nu_e \eta^{-6} de \sim \eta_{\rm
min}^{-6}$ for $e\to 1$, which is exactly the growth rate of the average number of peaks
given by $X(e_{\rm max})$.

In fact, to strictly ensure that no potential peak is missed, we should distribute the grid
nodes along all parametric isolines according to the corresponding pessimistic cases. We
violated this rule in the case of the eccentricity
isoline, for which the pessimistic density growth rate should be $\eta^{-2.5}$ instead
of the adopted $\eta^{-2}$. Otherwise we would have the total number of the grid nodes
growing as $\eta_{\rm min}^{-6.5}$, which is slightly larger than the total number
of the peaks given by $X(e_{\rm max})$. Such grid would be slighly oversampled in comarison
with an ideal (optimal) one. This oversampling appeared because the geometry
of the peaks may be distorted by the off-diagonal elements of $\mathbfss G$: a single peak
may become elongated and inclined, spanning across several grid layers, being counted in
each. To equate the number of grid nodes with the number of the expected peaks, we
neglected the multiplier of $1/\sqrt{\eta}$ in the eccentrictity density function. Even for
$e=0.9$ we have $\sqrt\eta\approx 0.66$, which does not differ much from unity. In practice
the need to handle $e>0.9$ emerges only in very rare extreme cases.

Also, we do not take into account the non-uniform distribution of $\omega$. It follows from
the above discussion that the sampled values of $\omega$ should concentrate somewhat to the
values of $0$ and $\pi$. However, the required concentration looks rather weak. Assuming
that the grid density should vary proportionally to $\sqrt{V_{\omega\omega}}$ (the density
of the peaks found on the given $\omega$-isoline), we obtain that the cumulative number of
such nodes in a variable range $[0,\omega]$ should be proportional
to $\arctan(\tan\omega/\sqrt\eta)$. This means that the necessary grid nodes can
be generated as $\omega = \arctan(\sqrt\eta \tan\alpha)$, where $\alpha$ is an auxiliary
uniformly distributed angle. Due to the same rather mild factor of $\sqrt\eta$, for $e<0.9$
this distribution does not differ very much from the uniform one.

Summarizing, our parametric grid can be constructed using the following instructions:
\begin{enumerate}
\item Sample the eccentricity $e$ according to the formula $e_k = \sqrt{1-\exp(-(2k+1)h_e)}$,
where $h_e$ controls the eccentricity resolution. This discrete distribution corresponds to
the eccentricity density function of $e/\eta^2$ (uniform in $\ln\eta$), satisfying the
$e\to 1$ asymptotic of $\sim \eta^{-2}$ requested above.
\item For each sampled value of $e$, construct the grid in $f$ with a step
of $h_f\eta^3/T$, where $h_f$ is a control parameter.
\item For each sampled value of $e$, construct the grid in $\lambda_0$ with a step
of $h_\lambda\eta^3$, where $h_\lambda$ is another control parameter.
\item For each sampled value of $e$, construct the grid in $\omega$ with some step
$h_\omega$.
\end{enumerate}
This grid has four control parameters $h_e,h_f,h_\lambda$, and $h_\omega$ that allow
to control the absolute resolution of the parameters. We recommend the following values:
$h_e=1/6$, $h_f=1/2$, $h_\lambda=\pi$, and $h_\omega=\pi/2$. Note that the
eccentricity should be sampled first, and the remaining Keplerian parameters
$f$, $\lambda_0$, and $\omega$ can be sampled independently from each other, but depending
on $e$.

It is rather obvious that the mean longitude $\lambda_0$, basically the phase of the signal,
should be sampled at higher density for large $e$: to locate the position of a narrow peak
in the high-eccentricity Keplerian curve (marking the periaston passage time) we should try
to fit many template Keplerian curves, each shifted by the width of the peak. However, the
similar property of the frequency might be surprising, although it might be suspected e.g.
from a similar property of the multiharmonic periodograms that require finer frequency
resolution with a larger order \citep{Baluev09a}. Consider that the model~(\ref{kepsig})
approximates the true signal in the middle of the observation segment, but the model
frequency is shifted from the true value by some $\Delta f$. Then near the ends of the time
segment the deviation between the phases of the model and of the true signal would be $\pm
\Delta f T/2$. This quantity should be at least the as the
step $\Delta \lambda_0$, because otherwise the periastron passages near the ends of the time
segment might displace too much, leading to an inadequate model.

After the grid is ready, we may run the Levenberg-Marquardt fitter per each grid
point, taking it as a starting approximation. The maximum likelihood attained over the grid
represents the desired value of the Keplerian periodogram. The associated best fitting
Keplerian parameters represent some useful by-product data.

A prototype algorithm of the Keplerian periodogram computation, based on the statistic
$\tilde Z$ from~(\ref{lrZ}), was included in the PlanetPack software \citep{Baluev13c}
as of version 1.6. In the forthcoming version~2.0 this algorithm will be released in the
improved form, including new optimized grid of Keplerian parameters described above.

\section{Keplerian periodogram in action}
\label{sec_demo}
As a good test suite for the Keplerian periodogram we consider the public RV data from
ELODIE \citep{Naef01},
Keck \citep{Butler06}, and HET \citep{Wittenmyer07b}\footnote{It appeared that
\citet{Wittenmyer09} released an improved version of the HET RV data for HD~80606. But
we discovered this already after our simulations were done, and we decided not to re-run
them, as we pursue only demonstration goals here.} for the famous star HD~80606, hosting a
unique planet that moves along an extremely elongated orbit ($e=0.93$).

\begin{figure}
\includegraphics[width=84mm]{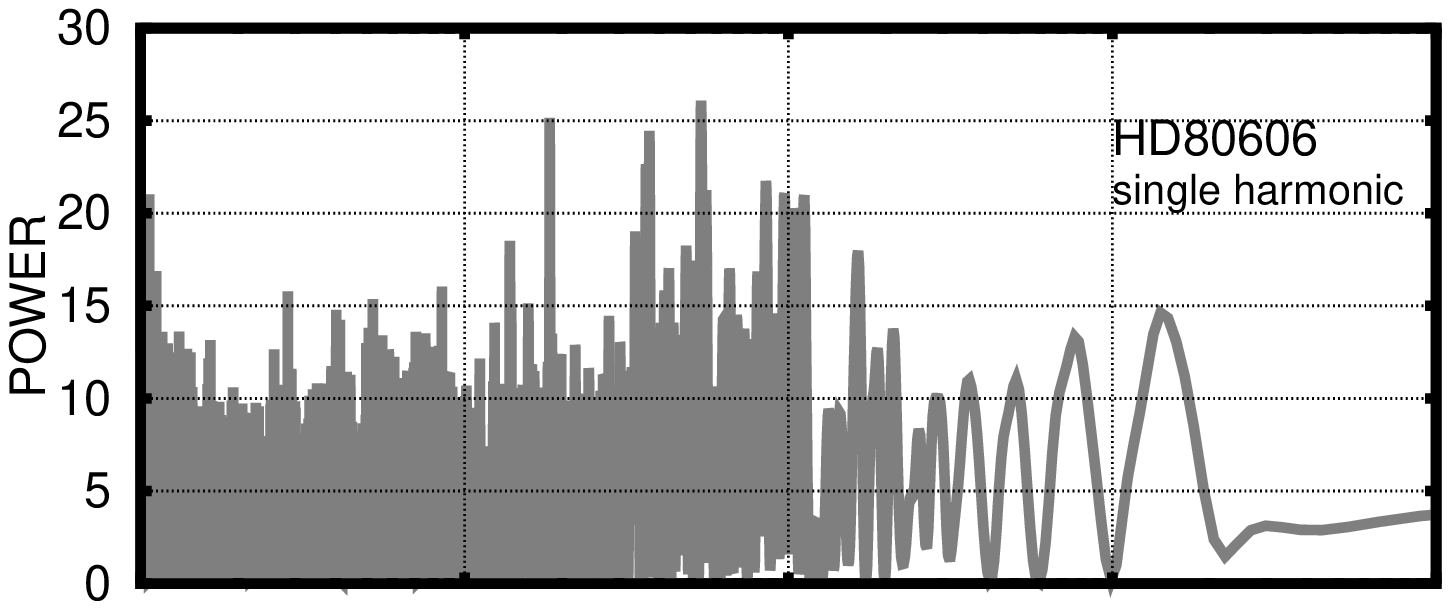}\\
\includegraphics[width=84mm]{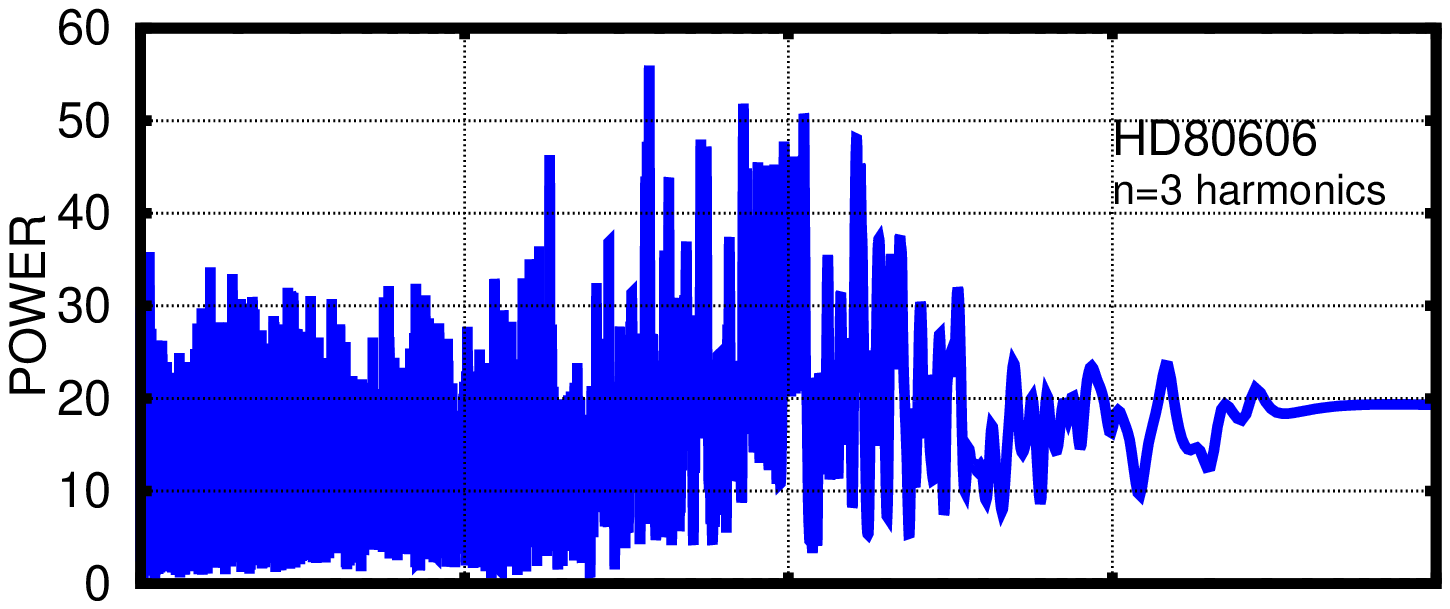}\\
\includegraphics[width=84mm]{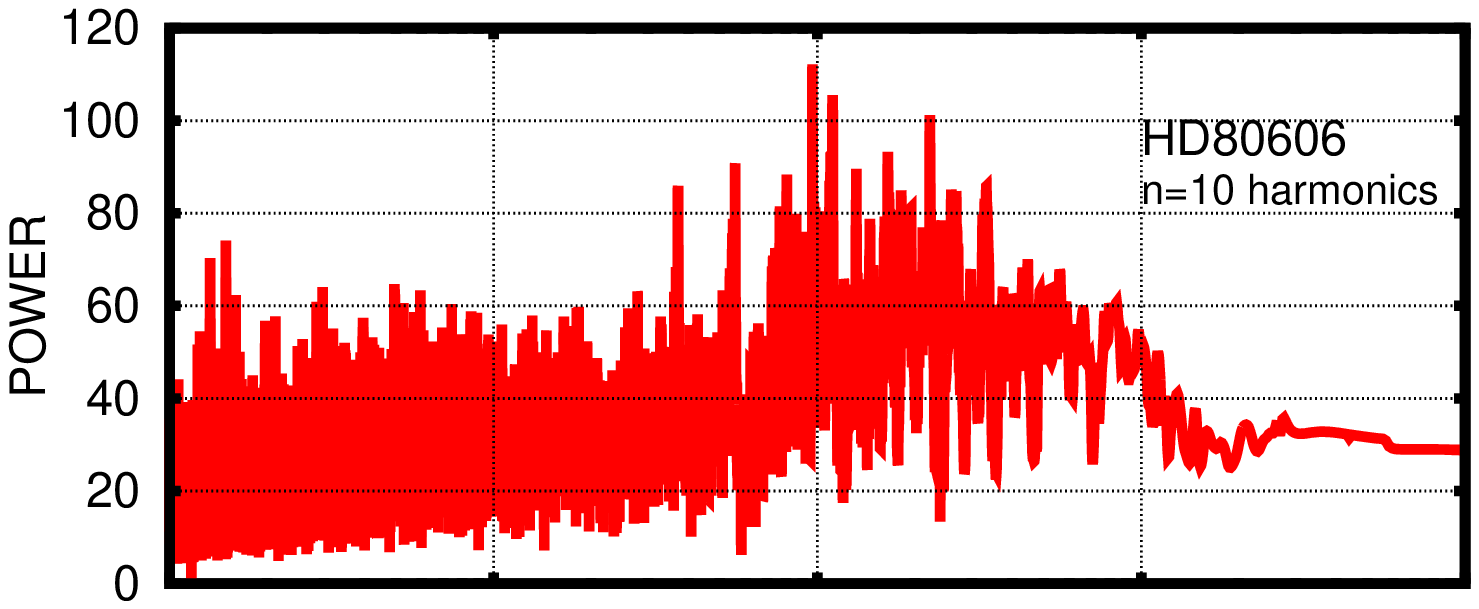}\\
\includegraphics[width=84mm]{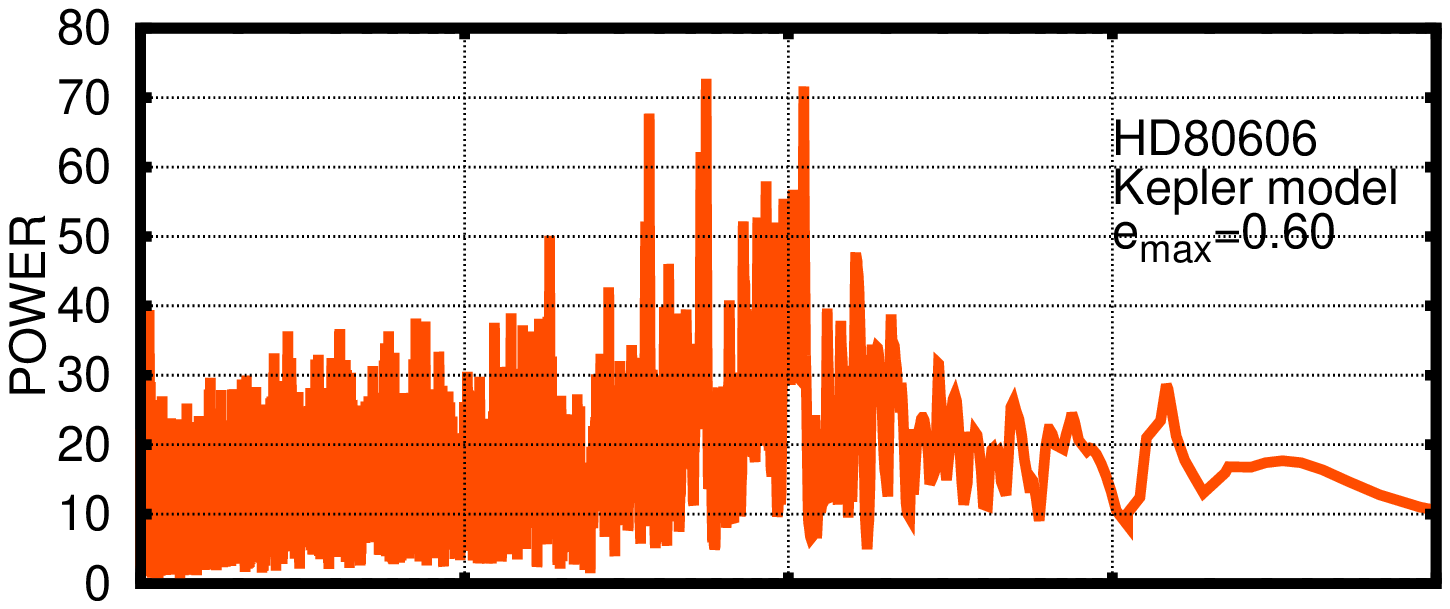}\\
\includegraphics[width=84mm]{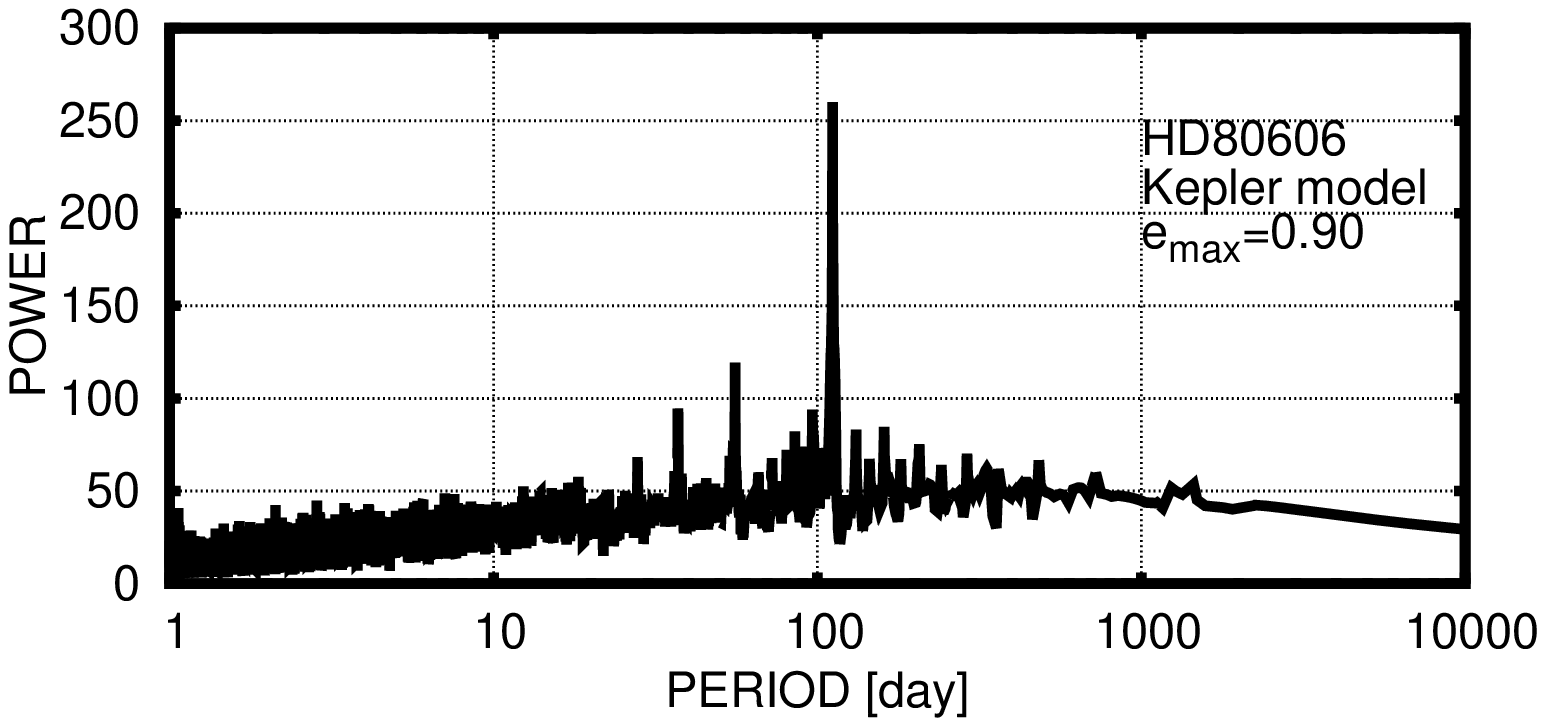}
\caption{Periodograms of the Doppler RV data for the star HD80606. From top to bottom: the
single-harmonic periodogram, two multiharmonic periodograms of different order, two
Keplerian periodograms with different maximum eccentricity.}
\label{fig_HD80606}
\end{figure}

First of all, we tried to process these data using more traditional sinusoidal model of
the signal. As expected, this periodogram did not releal anything distinguishable from the
noise (Fig.~\ref{fig_HD80606}, top frame), despite of a very large amplitude of the RV
variation. After that, we applied more advanced multiharmonic periodograms
\citep{SchwCzerny96,Baluev09a} to the data. In these periodograms the signal is modelled
as a sum of a few of first Fourier terms, which is more adapted to non-sinusoidal
variations. A disappointing thing is that even these periodograms do not help
(Fig.~\ref{fig_HD80606}, second and third plots). Even the model with $10$ Fourier
harmonics is in fact useless here: the periodogram still looks like a wide-band noise
without a hint of any clear isolated peak. Although we can see some moderate peak at the
true period $P=111$~d, there are a lot of other peaks at different periods. We would be
unable to identify the correct peak until we know the true period. At last, we proceed to
the Keplerian periodograms (Fig.~\ref{fig_HD80606}, fourth and fifth plots). We considered
to values of $e_{\rm max}$ here, $0.6$ and $0.9$. While the periodogram for $e_{\rm
max}=0.6$ still remains rather unimpressive, the periodogram for $e_{\rm max}=0.9$
contains a clear and undoubtful peak at the correct period value of $111$~d.

This test case emphasizes the importance of careful processing of large eccentricities. We
could not detect the planet until we reach the values of $e$ as large as $0.9$, and this
parameteric domain we must carefully sampled in order to not miss any maxima of the
likelihood function. This is what our computation algorithm is aimed on. Unfortunately,
dealing with large eccentricities $e>0.6$ requires quickly increasing computation
resources. However, we treat this as a necessary sacrifice.

\begin{figure}
\includegraphics[width=84mm]{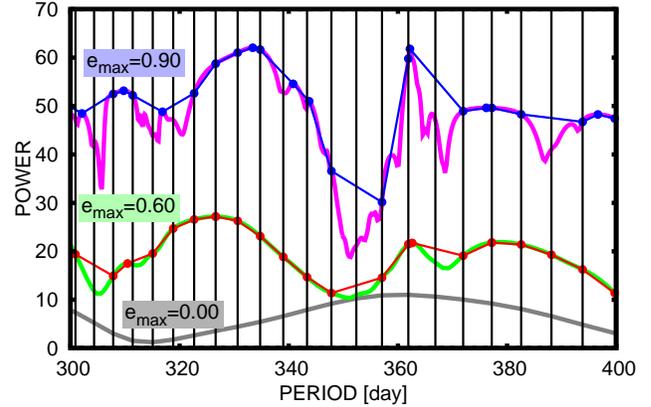}
\caption{The fine structure of the single-harmonic and Keplerian periodogram of the
HD80606 RV data. We plot short magnified pieces of two Keplerian periodograms from
Fig.~\ref{fig_HD80606} (points connected with line segments), and compare them with their
high-resolution versions (smooth lines). The grid of vertical lines maps to the frequency
grid adopted in Fig.~\ref{fig_HD80606}.}
\label{fig_HD80606mag}
\end{figure}

Now, let us consider how our algorithm works in more fine detailes. In
Fig.~\ref{fig_HD80606mag} we plot a zoomed graph of the single-harmonic periodogram and the
Keplerian periodograms for $e_{\rm max}=0.6$ and $e_{\rm max}=0.9$. Note that to reduce the
size of the output file, our Keplerian periodogram algorithm also allows to choose
an arbitrarily large effective (or userspace) frequency step, keeping internally the
fine grid necessary to adequately cover all likelihood maxima. The
user can set the frequency step even much larger than the typical width of the periodogram
peaks. The Keplerian periodogram is internally maximized in each of the large chunks of the
user-specified frequency grid, and only the maxima attained within the chunks are saved in
the file. We call this a ``sparse'' algorithm of periodogram evaluation. Two Keplerian
periodograms shown in Fig.~\ref{fig_HD80606mag} are plotted for two user-specified
resolutions: a version with very fine resolution (smooth curves) and a version
with low resolution (points connected by line segments). We also
overplot the frequency grid inferred by the latter (low) resolution.

First, we can see that $e_{\rm max}=0.9$ indeed generates much finer structures in the
periodogram than $e_{\rm max}=0.6$, as follows from the discussion in
Sect.~\ref{sec_comp}. If we would try to obtain the $e_{\rm max}=0.9$ periodogram fully
resolved and in the entire frequency segment of Fig.~\ref{fig_HD80606}, we would deal with
a huge output file and an increased computation time. Although we sacrificed the density
of frequency grid, we did not loose any important information about the periodogram peaks.
Each point of the output grid corresponds to either a local maxima inside a grid chunk, or
to a boundary of the chunk, depending on which periodogram value appears larger. Thus,
armed with this algorithm, we never loose any of the major peaks on the Keplerian
periodogram. In the ultimate case, we may even request to cover the entire frequency range
by a single step. Then the algorithm will just compute the absolute maximum of the
Keplerian periodogram attained over this frequency segment. The default frequency
resolution adopted in PlanetPack is such that approximately a single output value is
supplied per each local maximum of the plain single-frequency periodogram. This means that
in Fig.~\ref{fig_HD80606mag} we would have only a single such point, corresponding to the
maximum found within this range.

\section{Investigating the detection efficiency of the Keplerian periodogram}
\label{sec_eff}
In this section we undertake an attempt to quantitively characterize the detection
efficiency of the Keplerian periodogram. This is a rather rough investigation. We do not
try to deal with probabilistic efficiency characteristics like e.g. the probability of
planet detection for given Keplerian parameters. Instead, we only operate with the expected
detection thresholds, roughly corresponding to some intermediary (e.g. median) detection
probability. Also, we neglect the aliasing effects (or spectral leakage), which are caused
by an interference between the periodic gaps in the RV data and the periods of the
RV variation. We adopt the UPC (Uniform Phase Coverage) approximation noted above, in which
the averages over the time series are approximated by continuous integrals. Our goal
here is to characterize the detection efficiency of the Keplerian periodogram from the most
general point of view, instead of e.g. binding to the data with particular characteristics.

The power of a sinusoidal signal, with a given amplitude of $K$, is determined trivially
through the integration along its single period:
\begin{eqnarray}
\mathcal P_{\rm sin}(K) = \int\limits_0^{P} K^2 \cos^2\left(2\pi\frac{t}{P}+\lambda_0\right) \frac{dt}{P} = \nonumber\\
 = \int\limits_0^{2\pi} K^2 \cos^2\lambda\, \frac{d\lambda}{2\pi} = \frac{K^2}{2}.
\label{powsin}
\end{eqnarray}
The power of the Keplerian signal~(\ref{kepsig}) can be expressed in a similar way as
\begin{eqnarray}
\mathcal P_{\rm Kep}(K,\beta,\omega) = \int\limits_0^P \mu^2(t,\btheta) \frac{dt}{P} = K^2 \int\limits_0^{2\pi} g^2(\lambda,\bnu) \frac{d\lambda}{2\pi} = \nonumber\\
 = \frac{K^2}{2} \left(\frac{1-\beta^2}{1+\beta^2}\right)^2 \left(1-\beta^2\cos2\omega\right).
\label{powkep}
\end{eqnarray}
This result can be also found in the attached MAPLE worksheet.

In what follows below we largerly rely on the assumption that whenever the RV data carry a
signal indeed, the periodograms maxima should be approximately proportional to the power of
this signal, computed according to~(\ref{powsin}) or~(\ref{powkep}).

This property is easier to demonstrate for the simplest $\chi^2$ periodograms designated
above as $z(f)$ this property. First, we ``denoise'' this $\chi^2$
function~(\ref{chisq}) by replacing the measurements $x$ by the actual variation
(containing the real signal), and after that apply the UPC approximation:
\begin{eqnarray}
\chi^2_{\mathcal H,\mathcal K} = \left\langle (x-\mu_{\mathcal H,\mathcal K})^2 \right\rangle \simeq
 \left\langle (\hat\mu_{\mathcal K}-\mu_{\mathcal H,\mathcal K})^2 \right\rangle \simeq\nonumber\\
 \simeq \langle 1\rangle \int\limits_{t_{\rm min}}^{t_{\rm max}} (\hat\mu_{\mathcal K}-\mu_{\mathcal H,\mathcal K})^2 \frac{dt}{T}, \quad T=t_{\rm max}-t_{\rm min}.
\label{chisqUPC}
\end{eqnarray}
Here the notation $\hat\mu_{\mathcal K}$ stands for the actually present cumulative variation, having
the same functional shape as $\mu_{\mathcal K}$ (i.e. underlying variation + signal)
with some adopted ``true'' values of the parameters.

To derive the maximum of the relevant periodogram we need to
minimize the functions $\chi^2_{\mathcal H,\mathcal K}$ by varying the arguments
$\btheta_{\mathcal H,\mathcal K}$ in the models $\mu_{\mathcal H,\mathcal K}$, but keeping
the analogous parameters in $\hat\mu_{\mathcal K}$ fixed
at their prescripted values. Clearly, in the approximation~(\ref{chisqUPC}), the
global minimum for $\chi^2_{\mathcal K}$ is zero, achieved when the parameters
in $\mu_{\mathcal K}$ coincide with those in $\hat\mu_{\mathcal K}$. To handle
$\chi^2_{\mathcal H}$, we can apply a linearization of the base model $\mu_{\mathcal
H}$, if this model is not already linear in itself:
\begin{eqnarray}
\chi^2_{\mathcal H} \simeq \langle 1\rangle \int\limits_{t_{\rm min}}^{t_{\rm max}} (\hat\mu_{\mathcal H}-\mu_{\mathcal H}+\hat\mu)^2 \frac{dt}{T}.
\label{chisqUPC2}
\end{eqnarray}
Now, if the terms of the type
\begin{equation}
\int\limits_{t_{\rm min}}^{t_{\rm max}} \mu(t,\btheta) \mu_{\mathcal H}(t,\btheta_{\mathcal H}) dt
\label{orth}
\end{equation}
can be neglected then the integral in~(\ref{chisqUPC2}) can be split
in two independent nonnegative terms as
\begin{eqnarray}
\chi^2_{\mathcal H} \simeq \langle 1\rangle \left[ \int\limits_{t_{\rm min}}^{t_{\rm max}} (\hat\mu_{\mathcal H}-\mu_{\mathcal H})^2 \frac{dt}{T} + \int\limits_{t_{\rm min}}^{t_{\rm max}} \hat\mu^2 \frac{dt}{T} \right],
\end{eqnarray}
and then the minimum of $\chi^2_{\mathcal H}$ is achieved for $\btheta_{\mathcal
H} = \hat\btheta_{\mathcal H}$, when $\mu_{\mathcal H}$ coincides with
$\hat\mu_{\mathcal H}$. In this case we can easily obtain the desired result:
\begin{equation}
z = \frac{1}{2}(\min\chi^2_{\mathcal H}-\min\chi^2_{\mathcal K}) \simeq \frac{\langle 1\rangle}{2} \int\limits_{t_{\rm min}}^{t_{\rm max}} \hat\mu^2 \frac{dt}{T} \simeq \frac{\langle 1\rangle}{2} \int\limits_0^P \hat\mu^2 \frac{dt}{P}.
\label{chisqpow}
\end{equation}
As the time range $T$ is usually large, we have replaced in the last formula the
integration along the entire segment $[t_{\rm min},t_{\rm max}]$ by an integral along only
a single period of the signal. The last integral represents the power of the signal,
like~(\ref{powsin}) or~(\ref{powkep}).

The condition that~(\ref{orth}) should be negligible is an orthogonality condition between
the signal and the base models. In our assumptions, and for a sinusoidal
or a Keplerian signal, it is usually satisfied. For example, frequently $\mu_{\mathcal H}$
contains only a constant, and in this case we only need the signal to be properly centred,
satisfying $\int_0^P \mu\, dt = 0$. For a linear or quadratic underlying
variation, the orthogonality condition is also approximately fulfilled, if the
time range covers many periods of the signal (i.e. this period is short in comparison with
$T$), and we neglect aliasing effects. Similarly, whenever any previously detected planets
persist in $\mu_{\mathcal H}$, the orthogonality condition is approximately satisfied,
unless periods of some of these planets are close to the period of the signal, which is an
impractical case, or they settle an interference with the signal via the
aliasing mechanism, which we neglect here.

When the existing signal is Keplerian, and with the use of the Keplerian model, we may
accumulate the full Keplerian power~(\ref{powkep}) in the observed periodogram
maximum. Using~(\ref{chisqpow}) we may approximate this maximum as
\begin{equation}
z_{\rm Kep} \simeq \frac{\langle 1 \rangle}{2} P_{\rm Kep}(K,\beta,\omega).
\label{Zkep}
\end{equation}
Note that this approximation is valid only if the signal eccentricity does not
exceed the maximum one allowed in the computation of the Keplerian periodogram.

But if we use a sinusoidal model to detect a Keplerian signal we would deal with
multiple periodogram peaks corresponding to various Fourier subharmonics of~(\ref{kepsig}).
The height of each such peak would be proportional to the power of the relevant sinusoidal
subharmonic:
\begin{equation}
z_{\rm sin} \simeq \frac{\langle 1 \rangle}{2} P_{\rm sin}(K A_{\rm max}(\beta,\omega)),
\label{Zsin}
\end{equation}
where $A_{\rm max}$ is the maximum amplitude among the Fourier
subharmonics for the normalized Keplerian function $g(\lambda,\bnu)$.\footnote{Note that
for large eccentricities, the primary subharmonic is not necessarily the one
with the maximum amplitude.}

Thus, from~(\ref{Zkep}) and~(\ref{Zsin}) we can approximate the maxima ratio for the
Keplerian and a sinusoidal (e.g. Lomb-Scargle) periodograms like:
\begin{equation}
\frac{z_{\rm Kep}}{z_{\rm sin}} \simeq \left(\frac{1-\beta^2}{1+\beta^2}\right)^2 \frac{\left(1-\beta^2\cos2\omega\right)}{A_{\rm max}^2(\beta,\omega)}.
\label{Zrat}
\end{equation}
This ratio does not depend on $K$.

To compute the quantity $A_{\rm max}$ in~(\ref{Zsin}) and~(\ref{Zrat}), we
must use the Fourier coefficient of the Keplerian RV function, that become pretty easy
to compute using the formulae of the mentioned above integrals of $r^n \cos k\upsilon$ in
\citep{Kholsh-twobody}. Designating the cosine Fourier coefficients as $c_k$, and the sine
coefficients as $s_k$, we may obtain:
\begin{equation}
c_k = 2 \frac{1-e^2}{e} J_k(ke) \cos\omega, \quad s_k = 2\eta J_k'(ke) \sin\omega,
\end{equation}
where $J_k$ are Bessel functions. Finally, the desired quantity $A_{\rm max}$ can be
computed as:
\begin{equation}
A_{\rm max}^2 = \max_{k\geq 1} (c_k^2+s_k^2).
\end{equation}

From the other side, based on the $\FAP$ formulae above,
we can compute the approximate detection thresholds for the relevant periodograms, given
some small critical $\FAP=\alpha$:
\begin{equation}
z_{\rm sin}^{\rm thr}(\alpha,W), \qquad z_{\rm Kep}^{\rm thr}(\alpha,W,\beta_{\rm max}).
\end{equation}

The maximum peak observed in the Keplerian periodogram is larger than the maximum of the
sinusoidal periodogram, because the Keplerian periodogram is able to accumulate the full
power of the signal. But from the other side, the noise level in the Keplerian periodogram
is higher than in the sinusoidal one, because the Keplerian model has more free parameters,
and also because the expected number of the noisy peaks in the Keplerian
likelihood function grows quickly when $\beta_{\rm max}$ increases. The main
question is: which tendency wins, depending on the signal's $\beta$ and $\omega$?

So far we did not put any constraints on the signal amplitude $K$. Now, let us assume that
the signal amplitude $K$ is such that for the sinusoidal periodogram we have a
boundary detection:
\begin{equation}
z_{\rm sin}/z_{\rm sin}^{\rm thr} = 1.
\end{equation}
With this prerequisite, for the Keplerian periodogram of the same signal we can easily
compute the analogous detection efficiency ratio:
\begin{equation}
\frac{z_{\rm Kep}}{z_{\rm Kep}^{\rm thr}} = \frac{z_{\rm sin}^{\rm thr}}{z_{\rm Kep}^{\rm thr}} \frac{z_{\rm Kep}}{z_{\rm sin}}.
\label{Kepeff}
\end{equation}
In fact, the quantity~(\ref{Kepeff}) represents the relative detection efficiency of the
Keplerian periodogram comparatively to the ones utilizing a sinusoidal signal model.
Selecting some reasonable values for $\alpha$, $W$, and $\beta_{\rm max}$,
we can numerically compute the threshold ratio
in~(\ref{Kepeff}). The remaining ratio $z_{\rm Kep}/z_{\rm sin}$ is approximated
in~(\ref{Zrat}) through a function of $\beta$ and $\omega$. Therefore, the detection
efficiency~(\ref{Kepeff}) can be viewed as a function of a location in the $(\beta,\omega)$
plane. Values of~(\ref{Kepeff}) exceeding unit indicate the expected advantage
of the Keplerian periodogram, while values below unit indicate the advantage
of the sinusoidal model. The square root of this function measures the relative efficiency
in terms of the signal amplitude $K$ (rather than $K^2$, which is a less intuitive
quantity).

\begin{figure*}
\includegraphics[width=0.99\textwidth]{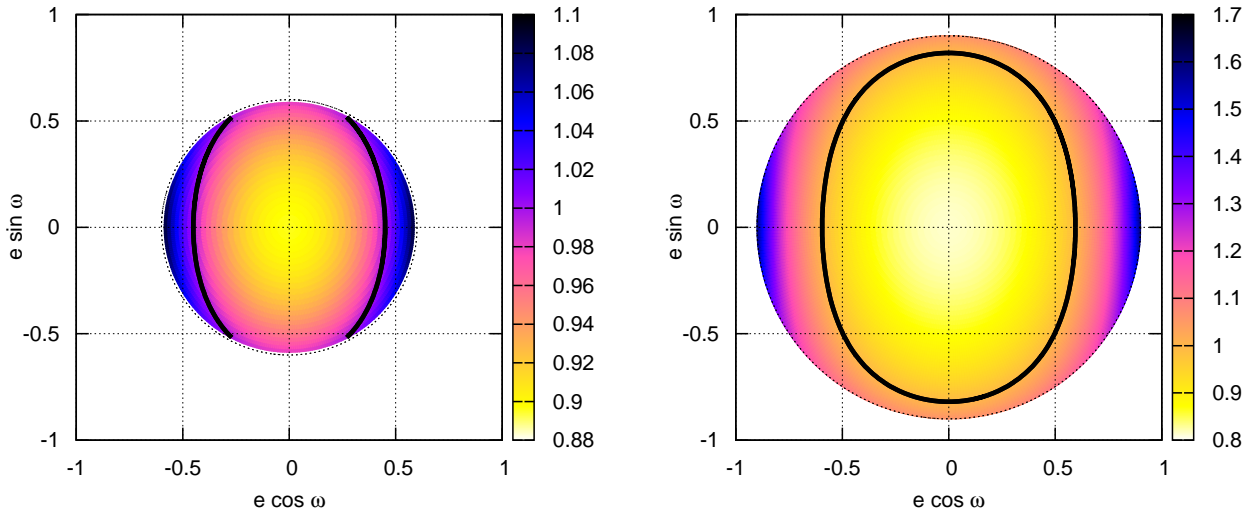}
\caption{Relative detection efficiency of the Keplerian periodogram comparatively
to the Lomb-Scargle one, plotted as a color-mapped function of the
planet orbital eccentricity $e$ and the pericenter argument $\omega$. The circular outline
of the plots labels the value of $e_{\rm max}$ adopted
in the Keplerian periodogram ($e_{\rm max}=0.6$ in the left panel and $0.9$ in the right
one). The oval thick line labels the points of equal detection efficiency (unit relative
efficiency). In the both plots we adopted the $\FAP$ threshold of $\alpha=0.01$, and the
frequency bandwidth of $W=5000$. See text for more comments and definitions.}
\label{fig_detect}
\end{figure*}

This relative detection efficiency is plotted in Fig.~\ref{fig_detect} as a function of the
planetary eccentric parameters $e\cos\omega$ and $e\sin\omega$. The two plots assume two
values of $e_{\rm max}$, respectively $0.6$ and $0.9$. The
$\FAP$ detection threshold was set to $\alpha=0.01$ and the frequency range to
$W=5000$ (other reasonable values did not lead to any remarkable changes
in the plots). First, we can see from these plots that they do not reveal any dramatic
difference in the detection efficiency. In the case $e_{\rm max}=0.6$ the relative
efficiency always stays close to unit. For $e_{\rm max}=0.9$, the minimum efficiency of the
Keplerian periodogram does not fall below $0.8$ of that of the Lomb-Scargle one.
As expected, this minimum corresponds to planets with zero eccentricities, for which the
use of the Keplerian model is unnecessary. The maximum relative efficiency reaches a
moderate value of $1.7$. corresponds to the points $\omega=0$ or $\omega=\pi$ located on
the boundary $e=e_{\rm max}$. In fact, for larger $e$ its relative efficiency should
grow further, but in the plots we cut out these regions, because the formula~(\ref{powkep})
becomes invalid there and we are unable to correctly compute the efficiency function there.

One might draw a conclusion from Fig.~\ref{fig_detect}
that the Keplerian periodogram can advance the detection power of highly eccentric planets
only pretty moderately, although for the unfavoured almost circular orbits it does not
introduce a significant degradation as well. However, these results were
obtained using a pretty rough and simplified treatment, and such a conclusion would not be
entirely objective. The practical advantage of the Keplerian periodogram for large
eccentricities does not relies on only the increased height of the periodogram peaks. In
Fig.~\ref{fig_HD80606}, top panel, we can see that the most important issue with the
sinusoidal periodogram is that it is unable to reveal a well isolated period
of the eccentric planet. This periodogram still looks like a wide-band noise. Formally, its
noise level is larger than we would expect from the data with no signal at all: e.g. the
estimation~(\ref{lsFAP}) yields $\FAP\sim 10^{-7}$. However,
this information remains rather useless, because we are unable to locate a clear period and
even to suspect that such a period exists. The Keplerian periodogram solves this
task gracefully, and not just because the height of its peaks is plainly larger. It allows
to locate a clearly isolated Keplerian period, indicating that the data indeed contain a
periodic signal. Moreover, from the results by \citet{OToole09b} it follows that the
Keplerian periodogram may become pretty useful even for moderate eccentricities of
$e\sim 0.6$. In this example it helped to disentangle multiple Keplerian signals from each
other, which the sinusoidal periodograms were unable to achieve.

In fact, the main purpose of Fig.~\ref{fig_detect} here is to demonstrate that when we are
dealing with planets having \emph{small} eccentricities, for which our efficiency indicator
is more adequate, the Keplerian periodogram still does not cause a significant degradation
in the detection power. This means that the Keplerian periodogram does not
impose any difficult trade-off between the detection of only low-eccentricity or
only high-eccentricity planets. This property becomes important for systems that contain
planets on orbits with low and high eccentricities simultaneously:
we could observe the hints of all such planets in the same Keplerian
periodogram. The sinusoidal periodogram would, at the best, reveal only the periods of the
low-eccentricity planets, presenting the other planets as a noisy mesh of peaks.

\section{Practical validity of the Keplerian significance thresholds}
\label{sec_simul}
The analytic $\FAP$ estimates from Sect.~\ref{sec_fap} were derived for pretty
simplified and apparently restrictive conditions. They are summarized here:
\begin{enumerate}
\item The simplified chi-square objective function~(\ref{chisq}) is adopted, implying that
the periodogram are based on the chi-square statistic $z$. This infers an assumption
of fixed and a priori known noise variances. In practice we usually do not
know the noise level well, implying that we have to use some parametrized noise model. This
would make use of the likelihood function~(\ref{likmod}) and likelihood-ratio test
statistic $Z$ or $\tilde Z$, involving e.g. an additive jitter model. This
is different from what we adopted in Sect.~\ref{sec_fap}.
\item The base model $\mu_{\mathcal H}$ is assumed strictly linear. In practice this
is true only if we have not detected even a single planet yet, implying that $\mu_{\mathcal
H}$ only involves a free RV offset, constant in time. But after the detection of the first
planet, the model $\mu_{\mathcal H}$ becomes non-linear. Even if this planet had
zero eccentricity and its RV signal was sinusoidal, at least the period of the
sinusoid would be a non-linear parameter.
\item To obtain entirely analytic $\FAP$ estimations in a closed form, the approximation of
the ``uniform phase coverage'' (UPC) was used extensively. This means that various
summations over the time series were approximated by continuous integrals over a
single period of $\mu$. In practice this approximation may be bad for uneven
time series, in particular when the signal period is in a commensurability with some
periodic leaks in the RV data. Also, this approximation may fail when the function to be
integrated contains narrow spikes or peaks that may fall in the gaps between
discrete observations. Such effect can emerge for large eccentricities.
\end{enumerate}
Fortunately, a violation of these assumptions does not necessarily corrupt the accuracy of
the final $\FAP$ estimations very much. Here is the justification:
\begin{enumerate}
\item Various statistical properties of the chi-square test are often applicable
to the likelihood-ratio one in an approximate (asymptotic) sense, under an extra condition
$N\to\infty$. This issue is considered in \citep{Baluev14a}, where it is also demonstrated
that the effect of a non-trivial noise model is similar in its nature to the effect of
non-linearity in the RV curve models, $\mu_{\mathcal H}$ in our case. Often this model can
be linearized in a small ($\sim 1/\sqrt N$) vicinity of the best fit parameters, again
making the results of Sect.~\ref{sec_fap} applicable approximately for large $N$. More
accurately, we should satisfy the asymptotic condition of the type $Z,\tilde Z \ll \mathcal
O(N)$ \citep{Baluev08b}. The results of Sect.~\ref{sec_fap} have asymptotic nature
themselves, requiring $z\to\infty$. Therefore, we must have $\tilde Z$ large enough to have
the formulae from Sect.~\ref{sec_fap} useful, but $\tilde Z$ must not be too much large, in
order to avoid the non-linearity effects. In practice it is enough to have good accuracy in
a rather limited range $\FAP \in [10^{-1}-10^{-3}]$: for small $\tilde Z$ levels the $\FAP$
appears too large anyway, implying an insignificant signal, and for very large $\tilde
Z$ it is anyway safely small (not a big deal, how much small).
\item As explained in \citep{Baluev13b}, the spectral leakage has a negligible effect
on the accuracy of the UPC approach, because the UPC approxaimation is invalidated only in
a few of very narrow frequency segments, associated to the peaks of the spectral
window function, while the necessary integrals usually involve a wide frequency range.
On contrary, the issue with failing UPC at large eccentricities may be potentially
important, although this effect is unrelated to the spectral leakage. When UPC
is failed, the coefficients $X$ and $Y$ in~(\ref{kfap}) should be computed using
direct time-series summations \citep[sect.~4.2]{Baluev13b}. We have little to simplify
or detail here: we should just substitute the parametric derivatives of the Keplerian RV
model~(\ref{kepsig}) and of the adopted $\mu_{\mathcal H}$ in that formulae. Although more
general and accurate, this method should be avoided whenever possible, because it is
computationally expensive (its complexity is comparable to a single evaluation
of the Keplerian periodogram), and the result is useful only for a particular time series.
\end{enumerate}

As previous simulations revealed \citep{BaluevBeauge14,Baluev14a}, the effect of non-linear
$\mu_{\mathcal H}$ as well as the effect of the non-trivial noise model may increase $\FAP$
in comparison with what expected from the linear case. This may even break the
inequality in~(\ref{fapbase}). However, the inaccuracy of the UPC approximation likely has
an opposite effect, leading to an overestimated $\FAP$ \citep{Baluev13b}. We do not expect
it to break the inequality of~(\ref{fapbase}). Whether or not these effects are significant
at all, depends on the particular case. In this section our goal is to
verify the applicability of the $\FAP$ estimations under various
conditions typical for exoplanetary Doppler surveys.

\begin{figure}
\includegraphics[width=84mm]{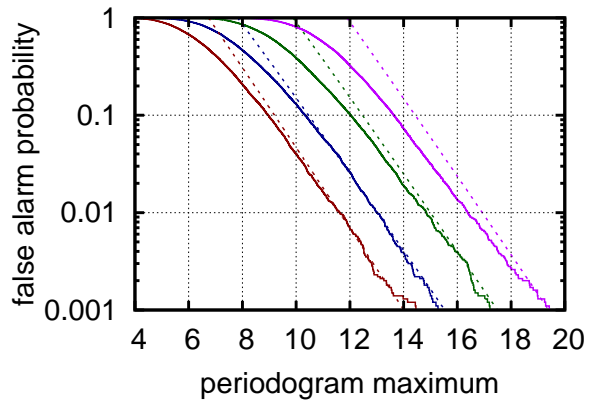}
\caption{Comparison of the simulated $\FAP$s of the Keplerian periodogram with their
analytic estimations: the case of the 51~Peg ELODIE RV data. The graphs show the simulated
$\FAP$s (solid curves) with their theoretic approximations (broken
lines) for $e_{\rm max}=0$, $0.3$, $0.6$, and $0.8$ (from left
to right). All the cases have $f_{\rm max}=0.1$~d$^{-1}$.}
\label{fig_51Peg_pow}
\end{figure}

In the first example we consider the 51~Pegasi RV data acquired by ELODIE spectrograph
\citep{Naef04}. Their time distribution is more or less typical for ground-based
surveys, involving seasonic gaps and diurnal regularity. Their number is relatively large,
$N=153$, and the base model is rather simple, involving only a single planet on almost
circular orbit (implying only a single sinusoidal RV term). Such simple model is very well
linearizable. Concerning the RV noise in this data, we approximate it with the usual model
involving an additive jitter \citep{Baluev08b,Wright05}. In another study we
have already verified that it does not generate any significant additional non-linearity
effects with these particular data \citep{Baluev14a}. When dealing
with a Keplerian periodogram of these data, we may face only the non-linearity of the probe
Keplerian signal~(\ref{kepsig}). This makes the task very close to the idealized conditions
of Sect.~\ref{sec_fap}.

Monte Carlo simulations presented in Fig.~\ref{fig_51Peg_pow} confirm this. We can see that
the simulated $\FAP$ curves are in a good agreement with the
analytic formulae~(\ref{kfap},\ref{XYapprox}). Monte Carlo simulation of the Keplerian
periodogram is an extremely heavy task in view of the computational resources, so we had to
limit the periodogram to a rather narrow frequency range
with $f_{\rm max}=0.1$~d$^{-1}$, although the more practical and typically adopted value is
$f_{\rm max}=1$~d$^{-1}$. We sacrificed the frequency range in order to allow large enough
values of the eccentricity limit $e_{\rm max}$, because the specific
of the Keplerian periodogram is in the variable eccentricity. In fact,
this further highlights the usefullness of the analytic $\FAP$ estimation for the Keplerian
periodogram: for more pratical values of $f_{\rm max}$ and $e_{\rm max}$ it is still
possible to compute a single Keplerian periodogram, but it would be just infeasible to
simulate its $\FAP$ levels by Monte Carlo.

\begin{figure}
\includegraphics[width=84mm]{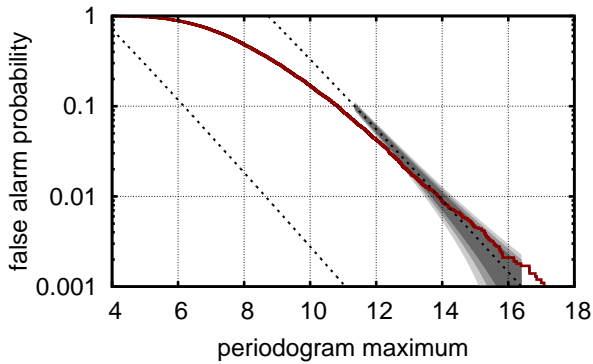}
\caption{Comparison of the simulated $\FAP$ of the Keplerian periodogram with the
analytic estimations: the case of the HD~80606 RV data. The simulated $\FAP$ for $e_{\rm
max}=0.9$ and $f_{\rm max}=0.01$~d$^{-1}$ is shown as a solid curve. The first broken line
(in the left part) corresponds to the analytic $\FAP$ for the Lomb-Scargle periodogram and
is shown here for comparison. The second broken curve shows the analytic approximation
to the Keplerian $\FAP$. The gray-filled areas around this line correspond
to the $1,2,3$-sigma Monte Carlo uncertainty ranges (see text for details).}
\label{fig_HD80606_pow}
\end{figure}

Another simulation refers to the RV data of HD~80606, already considered
in Sect.~\ref{sec_demo}. In this case we adopted more complicated base model, including the
high-eccentricity planet \emph{b}, independent offsets for the RV subsets coming
from different observatories, and an additional sinusoidal annual variation in the HET
RV data from \citep{Wittenmyer07b}. The latter variation was partly motivated by the
presence of a similar annual variation in some HET RV data for
HD~74156 \citep{Baluev08b,Meschiari11}. Although that HD~74156 data were acquired by
another team \citet{Bean08}, and we did not actually detect a significant annual variation
in the HD~80606 data, we added this annual term to the RV model to make it
more complicated. Finally, we replaced the additive RV noise model with the regularized one
from \citep{Baluev14a} to suppress possible interfering non-linearity generated by the
RV noise rather than RV curve.

The simulation results for this case are presented
in Fig.~\ref{fig_HD80606_pow}. The frequency limit here was reduced to $f_{\rm
max}=0.01$~d$^{-1}$, to allow the higher eccentricty limit of $e_{\rm max}=0.9$. We can see
that for large periodogram peaks, or equivalently small $\FAP$ levels, the simulated $\FAP$
curve passes slightly above the theoretic prediction, formally
breaking the inequality~(\ref{fapbase}). This may indicate the presence of some modest
non-linearity in the base RV model. In this case we may note that the impact
of non-linearity is smaller than we might expect. In the base model we have a $e=0.93$
Keplerian signal of the known planet, which is an extremely non-linear function from the
first view. Nonetheless, even this modest $\FAP$ increase is consistent with the Monte
Carlo uncertainties (shown in Fig.~\ref{fig_HD80606_pow} as gray-filled ranges
near the theoretically predicted curve). The latter uncertainties were constructed by means
of a tail-weighted Kolmogorov-Smirnov test
\citep{ChichBouch12,Baluev14a}.

\section{Conclusions and discussion}
Although a lot of period finding methods are currently available for an astronomer
\citep{Graham13}, it is important to avoid an unjustified use of a method in the tasks
in which it was not supposed to properly work. For example, the old approach
\citep{HorneBal86} approximates the periodogram $\FAP$ levels on the basis
of the assumption of independent periodogram readouts. For the
Lomb-Scargle periodogram this approach approximates the $\FAP$ with a formula:
\begin{equation}
\FAP \approx 1 - (1 - e^{-z})^{N_{\rm ind}},
\label{HB}
\end{equation}
where $N_{\rm ind}$ is the effective number of independent periodogram
readouts (or ``independent frequencies''). The formula~(\ref{HB}) is formally valid only
under pretty strict conditions, namely (i) the time series should be
evenly spaced, and (ii) the periodogram of these data is computed only on a discrete and
rather sparse set of the fundamental frequencies (no oversampling). As these conditions are
rarely fulfilled in the astronomical pracice, \citet{HorneBal86} suggested
to use~(\ref{HB}) just as an extrapolating formula, in which the quantity $N_{\rm ind}$ is
treated as a free parameter fitted by Monte Carlo simulations. We do not criticize
here this idea itself, because it was rather reasonable and in fact the most
usable approach among those available in 1986. However, today we should remember that in
the \citet{HorneBal86} treatment the formula~(\ref{HB}) have lost its theoretical basis and
serves as just a parametric $\FAP$ fitting formula. Its accuracy is not guaranteed
in the individual practical cases.

Moreover, numerous later works have already revealed multiple theoretic as well as
practical weaknesses of the formula~(\ref{HB}), see e.g.
\citep{Koen90,Frescura08,Baluev08a,Suveges14}. For
example, for large $z$ the formula~(\ref{HB}) yields the approximation $\FAP \sim N_{\rm
ind} e^{-z}$, which is by the factor of $\sqrt z$ different from the correct
asymptotic $\FAP$ behaviour, given by~(\ref{lsFAP}). Although this $\sqrt z$
factor could be compensated by selecting a larger $N_{\rm ind}$, to achieve this we should
significantly increase the number of Monte Carlo trials to cover the smaller $\FAP$ levels
reliably. However, in this case the use of the formula~(\ref{HB}) becomes just
senseless, because in this case we could just estimate the $\FAP$ from these simulations
directly.

The modern $\FAP$ estimation that were originally introduced
in \citep{Baluev08a} for the Lomb-Scargle periodogram, and now extended to the Keplerian
signal model, are free from the issues of the \citet{HorneBal86} approach. Our technique
has a strict and general theoretic basis of the Rice method, while its final
formulae are usually very simple and do not require any Monte Carlo simulations at
all. Moreover, this entirely analytic approximation in practice usually appears more
accurate than the approximation~(\ref{HB}), even if we fit $N_{\rm ind}$ in the latter
by simulations. For example, \citet{Hartman14} says that with the \citet{HorneBal86} method
``the resulting false alarm probability may be inaccurate by as much as a factor
of $\sim 10$'', and this seems to be even a rather optimistic assessment. In view of this,
it appears rather strange and unexplainable that an almost 30-years age method, which is
already known for its practical deficiency, is still so frequently used in the astronomical
practice \citep[e.g.][among publications of only the year of this
writing]{Burt14,Brothwell14,Kelly14,Hartman14,Nucita14,Chen14,Lin14,Romano14,Buccino14,Brucalassi14}.

It might be indeed a bit scaring to blindly rely on an entirely analytic formula
like~(\ref{lsFAP}) without any Monte Carlo calibration. However, instead of fitting the
archaic fromula~(\ref{HB}), which is already known to be not accurate, it might be
considerably more informative to just give a comparison of the simulated periodogram
distribution with~(\ref{lsFAP}) and~(\ref{HB}). Alternatively, it might be recommended to
fit the $\FAP$ with some more suitable approximation, e.g. with the two-parametric formula
\begin{equation}
\FAP \approx A e^{-z} z^p,
\end{equation}
with $A$ and $p$ determined by Monte Carlo. This formula represents a general form for the
primary $\FAP$ term that is usually obtained after applying the Rice method to various
periodograms. Also, \citet{Suveges14} suggested to fit the $\FAP$ with the three-parametric
generalized extreme-value distribution. As well as the Rice method, this new approach has a
general mathematical basis too, although it requires Monte Carlo, and the limits
of its applicability and accuracy look different.

In view of the increased complexity of the signal model used in the Keplerian periodogram,
its practical application is a computationally heavy task. In this concern, analytic
methods that allow to avoid Monte Carlo simulations become especially precious. The main
result of this paper is the analytic
$\FAP$ approximation for the Keplerian periodogram given
by the formulae~(\ref{kfap}) and~(\ref{XYapprox}) of Sect.~\ref{sec_fap}. As we can see,
these formulae are in fact elementary and should be definitely helpful
in practical computations. These $\FAP$ estimations, together with the computation
algorithm of Sect.~\ref{sec_comp}, are now implemented in PlanetPack, a public open-source
software for Doppler time series analysis \citep{Baluev13c}. Note that
some computation algorithm for the Keplerian periodogram was included in PlanetPack~1.6 and
onwards, but that was only a preliminary experimental version without any $\FAP$
estimations. The new algorithms will be soon released with the forthcoming PlanetPack~2.0.

\section*{Acknowledgements}
This work was supported by the President grant for young scientists (MK-733.2014.2), by the
Russian Foundation for Basic Research (projects No. 12-02-31119 mol\_a and 14-02-92615
KO\_a), and by the programme of the Presidium of Russian Academy of Sciences
``Non-stationary phenomena in the objects of the Universe''. The PlanetPack C++ code
includes a snapshot of the C library for stellar limb-darkening models
from \citep{AbubGost13}, by the kind permission of the authors. I would like to express my
gratitude to the anonymous referee for providing insightful comments.

\bibliographystyle{mn2e}
\bibliography{kepler}

\begin{thebibliography}{}
\makeatletter
\relax
\def\mn@urlcharsother{\let\do\@makeother \do\$\do\&\do\#\do\^\do\_\do\%\do\~}
\def\mn@doi{\begingroup\mn@urlcharsother \@ifnextchar[{\mn@doi@}{\mn@doi@[]}}
\def\mn@doi@[#1]#2{\def\@tempa{#1}\ifx\@tempa\@empty
  \href{http://dx.doi.org/#2}{doi:#2}\else \href{http://dx.doi.org/#2}{#1}\fi
  \endgroup}
\def\mn@eprint#1#2{\mn@eprint@#1:#2::\@nil}
\def\mn@eprint@arXiv#1{\href{http://arxiv.org/abs/#1}{{\tt arXiv:#1}}}
\def\mn@eprint@dblp#1{\href{http://dblp.uni-trier.de/rec/bibtex/#1.xml}{dblp:#1}}
\def\mn@eprint@#1:#2:#3:#4\@nil{\def\@tempa {#1}\def\@tempb {#2}\def\@tempc
  {#3}\ifx \@tempc \@empty \let\@tempc\@tempb \let\@tempb\@tempa \fi \ifx
  \@tempb \@empty \def\@tempb{arXiv}\fi \@ifundefined
  {mn@eprint@\@tempb}{\@tempb:\@tempc}{\expandafter \expandafter \csname
  mn@eprint@\@tempb\endcsname \expandafter{\@tempc}}}

\bibitem[\protect\citeauthoryear{Abubekerov \& Gostev}{Abubekerov \&
  Gostev}{2013}]{AbubGost13}
Abubekerov M.~K.,  Gostev N.~Y.,  2013, MNRAS, 432, 2216

\bibitem[\protect\citeauthoryear{Anglada-Escud{\'e} \&
  Tuomi}{Anglada-Escud{\'e} \& Tuomi}{2012}]{Anglada-Escude12}
Anglada-Escud{\'e} G.,  Tuomi M.,  2012, A\&A, 548, A58

\bibitem[\protect\citeauthoryear{Baluev}{Baluev}{2008}]{Baluev08a}
Baluev R.~V.,  2008, MNRAS, 385, 1279

\bibitem[\protect\citeauthoryear{Baluev}{Baluev}{2009a}]{Baluev08b}
Baluev R.~V.,  2009a, MNRAS, 393, 969

\bibitem[\protect\citeauthoryear{Baluev}{Baluev}{2009b}]{Baluev09a}
Baluev R.~V.,  2009b, MNRAS, 395, 1541

\bibitem[\protect\citeauthoryear{Baluev}{Baluev}{2013a}]{Baluev13c}
Baluev R.~V.,  2013a, Astronomy \& Computing, 2, 18

\bibitem[\protect\citeauthoryear{Baluev}{Baluev}{2013b}]{Baluev13e}
Baluev R.~V.,  2013b, Astronomy \& Computing, 3-4, 50

\bibitem[\protect\citeauthoryear{Baluev}{Baluev}{2013c}]{Baluev13b}
Baluev R.~V.,  2013c, MNRAS, 431, 1167

\bibitem[\protect\citeauthoryear{Baluev}{Baluev}{2013d}]{Baluev13d}
Baluev R.~V.,  2013d, MNRAS, 436, 807

\bibitem[\protect\citeauthoryear{Baluev}{Baluev}{2014a}]{Baluev14c}
Baluev R.~V.,  2014a, Astophysics, 57, 434

\bibitem[\protect\citeauthoryear{Baluev}{Baluev}{2014b}]{Baluev14a}
Baluev R.~V.,  2014b, MNRAS, accepted, arXiv:1407.8482

\bibitem[\protect\citeauthoryear{Baluev \& Beaug{\'e}}{Baluev \&
  Beaug{\'e}}{2014}]{BaluevBeauge14}
Baluev R.~V.,  Beaug{\'e} C.,  2014, MNRAS, 439, 673

\bibitem[\protect\citeauthoryear{Bean, McArthur, Benedict  \& Armstrong}{Bean
  et~al.}{2008}]{Bean08}
Bean J.~L.,  McArthur B.~E.,  Benedict G.~F.,   Armstrong A.,  2008, ApJ, 672,
  1202

\bibitem[\protect\citeauthoryear{{Brothwell} et~al.,}{{Brothwell}
  et~al.}{2014}]{Brothwell14}
{Brothwell} R.~D.,  et~al., 2014, MNRAS, 440, 3392

\bibitem[\protect\citeauthoryear{{Brucalassi} et~al.,}{{Brucalassi}
  et~al.}{2014}]{Brucalassi14}
{Brucalassi} A.,  et~al., 2014, A\&A, 561, L9

\bibitem[\protect\citeauthoryear{{Buccino}, {Petrucci}, {Jofr{\'e}}  \&
  {Mauas}}{{Buccino} et~al.}{2014}]{Buccino14}
{Buccino} A.~P.,  {Petrucci} R.,  {Jofr{\'e}} E.,   {Mauas} P.~J.~D.,  2014,
  ApJ, 781, L9

\bibitem[\protect\citeauthoryear{Burt, Vogt, Butler, Hanson, Meschiari, Rivera,
  Henry  \& Laughlin}{Burt et~al.}{2014}]{Burt14}
Burt J.,  Vogt S.~S.,  Butler R.~P.,  Hanson R.,  Meschiari S.,  Rivera E.~J.,
  Henry G.~W.,   Laughlin G.,  2014, ApJ, in press, arXiv:1405.2929

\bibitem[\protect\citeauthoryear{Butler et~al.,}{Butler
  et~al.}{2006}]{Butler06}
Butler R.~P.,  et~al., 2006, ApJ, 646, 505

\bibitem[\protect\citeauthoryear{{Chen}, {Hu}, {Guo}  \& {Du}}{{Chen}
  et~al.}{2014}]{Chen14}
{Chen} X.,  {Hu} S.~M.,  {Guo} D.~F.,   {Du} J.~J.,  2014, Ap\&SS, 349, 909

\bibitem[\protect\citeauthoryear{Chicheportiche \& Bouchaud}{Chicheportiche \&
  Bouchaud}{2012}]{ChichBouch12}
Chicheportiche R.,  Bouchaud J.-P.,  2012, Phys Rev E, 86, 041115

\bibitem[\protect\citeauthoryear{Cumming}{Cumming}{2004}]{Cumming04}
Cumming A.,  2004, MNRAS, 354, 1165

\bibitem[\protect\citeauthoryear{Cumming}{Cumming}{2010}]{CummingStat}
Cumming A.,  2010, Statistical Distribution of Exoplanets.
pp 191--214, in \cite{ExoplanetsSeager}

\bibitem[\protect\citeauthoryear{Ferraz-Mello}{Ferraz-Mello}{1981}]{FerrazMello81}
Ferraz-Mello S.,  1981, AJ, 86, 619

\bibitem[\protect\citeauthoryear{Foster}{Foster}{1995}]{Foster95}
Foster G.,  1995, AJ, 109, 1889

\bibitem[\protect\citeauthoryear{Frescura, Engelbrecht  \& Frank}{Frescura
  et~al.}{2008}]{Frescura08}
Frescura F. A.~M.,  Engelbrecht C.~A.,   Frank B.~S.,  2008, MNRAS, 388, 1693

\bibitem[\protect\citeauthoryear{Graham, Drake, Djorgovski, Mahabal, Donalek,
  Duan  \& Maker}{Graham et~al.}{2013}]{Graham13}
Graham M.~J.,  Drake A.~J.,  Djorgovski S.~G.,  Mahabal A.~A.,  Donalek C.,
  Duan V.,   Maker A.,  2013, MNRAS, 434, 3423

\bibitem[\protect\citeauthoryear{Gregory}{Gregory}{2007a}]{Gregory07a}
Gregory P.~C.,  2007a, MNRAS, 374, 1321

\bibitem[\protect\citeauthoryear{Gregory}{Gregory}{2007b}]{Gregory07b}
Gregory P.~C.,  2007b, MNRAS, 381, 1607

\bibitem[\protect\citeauthoryear{{Hartman} et~al.,}{{Hartman}
  et~al.}{2014}]{Hartman14}
{Hartman} J.~D.,  et~al., 2014, AJ, 147, 128

\bibitem[\protect\citeauthoryear{Horne \& Baliunas}{Horne \&
  Baliunas}{1986}]{HorneBal86}
Horne J.~H.,  Baliunas S.~L.,  1986, ApJ, 302, 757

\bibitem[\protect\citeauthoryear{{Kelly}, {Becker}, {Sobolewska},
  {Siemiginowska}  \& {Uttley}}{{Kelly} et~al.}{2014}]{Kelly14}
{Kelly} B.~C.,  {Becker} A.~C.,  {Sobolewska} M.,  {Siemiginowska} A.,
  {Uttley} P.,  2014, ApJ, 788, 33

\bibitem[\protect\citeauthoryear{Kholshevnikov \& Titov}{Kholshevnikov \&
  Titov}{2007}]{Kholsh-twobody}
Kholshevnikov K.~V.,  Titov V.~B.,  2007, Two-body problem [in Russian].
St. Pet. Univ. Press, St Petersburg

\bibitem[\protect\citeauthoryear{Koen}{Koen}{1990}]{Koen90}
Koen C.,  1990, ApJ, 348, 700

\bibitem[\protect\citeauthoryear{{Lin}, {Ip}, {Lin}, {Yoshida}  \&
  {Cheng}}{{Lin} et~al.}{2014}]{Lin14}
{Lin} C.-H.,  {Ip} W.-H.,  {Lin} Z.-Y.,  {Yoshida} F.,   {Cheng} Y.-C.,  2014,
  Research in Astronomy and Astrophysics, 14, 311

\bibitem[\protect\citeauthoryear{Lomb}{Lomb}{1976}]{Lomb76}
Lomb N.~R.,  1976, Ap\&SS, 39, 447

\bibitem[\protect\citeauthoryear{Mayor \& Queloz}{Mayor \&
  Queloz}{1995}]{MayorQueloz95}
Mayor M.,  Queloz D.,  1995, Nature, 378, 355

\bibitem[\protect\citeauthoryear{Meschiari, Laughlin, Vogt, Butler, Rivera,
  Haghighipour  \& Jalowiczor}{Meschiari et~al.}{2011}]{Meschiari11}
Meschiari S.,  Laughlin G.,  Vogt S.~S.,  Butler R.~P.,  Rivera E.~J.,
  Haghighipour N.,   Jalowiczor P.,  2011, ApJ, 727, 117

\bibitem[\protect\citeauthoryear{Naef et~al.,}{Naef et~al.}{2001}]{Naef01}
Naef D.,  et~al., 2001, A\&A, 375, L27

\bibitem[\protect\citeauthoryear{Naef, Mayor, Beuzit, Perrier, Queloz, Sivan
  \& Udry}{Naef et~al.}{2004}]{Naef04}
Naef D.,  Mayor M.,  Beuzit J.~L.,  Perrier C.,  Queloz D.,  Sivan J.~P.,
  Udry S.,  2004, A\&A, 414, 351

\bibitem[\protect\citeauthoryear{{Nucita}, {Giordano}, {De Paolis}  \&
  {Ingrosso}}{{Nucita} et~al.}{2014}]{Nucita14}
{Nucita} A.~A.,  {Giordano} M.,  {De Paolis} F.,   {Ingrosso} G.,  2014, MNRAS,
  438, 2466

\bibitem[\protect\citeauthoryear{O'Toole et~al.,}{O'Toole
  et~al.}{2007}]{OToole07}
O'Toole S.~J.,  et~al., 2007, ApJ, 660, 1636

\bibitem[\protect\citeauthoryear{O'Toole, Tinney, Jones, Butler, Marcy, Carter
  \& Bailey}{O'Toole et~al.}{2009a}]{OToole09a}
O'Toole S.~J.,  Tinney C.~G.,  Jones H. R.~A.,  Butler R.~P.,  Marcy G.~W.,
  Carter B.,   Bailey J.,  2009a, MNRAS, 392, 641

\bibitem[\protect\citeauthoryear{O'Toole, Jones, Tinney, Butler, Marcy, Carter,
  Bailey  \& Wittenmyer}{O'Toole et~al.}{2009b}]{OToole09b}
O'Toole S.~J.,  Jones H. R.~A.,  Tinney C.~G.,  Butler R.~P.,  Marcy G.~W.,
  Carter B.,  Bailey J.,   Wittenmyer R.~A.,  2009b, ApJ, 701, 1732

\bibitem[\protect\citeauthoryear{P{\'a}l}{P{\'a}l}{2010}]{Pal10}
P{\'a}l A.,  2010, MNRAS, 409, 975

\bibitem[\protect\citeauthoryear{{Romano} et~al.,}{{Romano}
  et~al.}{2014}]{Romano14}
{Romano} P.,  et~al., 2014, A\&A, 562, A2

\bibitem[\protect\citeauthoryear{Scargle}{Scargle}{1982}]{Scargle82}
Scargle J.~D.,  1982, ApJ, 263, 835

\bibitem[\protect\citeauthoryear{Schneider}{Schneider}{1995}]{Schneider}
Schneider J.,  1995, The Extrasolar Planets Encyclopaedia, www.exoplanet.eu

\bibitem[\protect\citeauthoryear{Schwarzenberg-Czerny}{Schwarzenberg-Czerny}{1996}]{SchwCzerny96}
Schwarzenberg-Czerny A.,  1996, ApJ, 460, L107

\bibitem[\protect\citeauthoryear{Seager}{Seager}{2010}]{ExoplanetsSeager}
Seager S.,  ed. 2010, Exoplanets.
University of Arizona Press, Tucson

\bibitem[\protect\citeauthoryear{S{\"u}veges}{S{\"u}veges}{2014}]{Suveges14}
S{\"u}veges M.,  2014, MNRAS, 440, 2099

\bibitem[\protect\citeauthoryear{Tuomi \& Anglada-Escud{\'e}}{Tuomi \&
  Anglada-Escud{\'e}}{2013}]{Tuomi13}
Tuomi M.,  Anglada-Escud{\'e} G.,  2013, A\&A, 556, A111

\bibitem[\protect\citeauthoryear{Wittenmyer, Endl, Cochran  \&
  Levison}{Wittenmyer et~al.}{2007}]{Wittenmyer07b}
Wittenmyer R.~A.,  Endl M.,  Cochran W.~D.,   Levison H.~F.,  2007, AJ, 134,
  1276

\bibitem[\protect\citeauthoryear{Wittenmyer, Endl, Cochran, Levison  \&
  Henry}{Wittenmyer et~al.}{2009}]{Wittenmyer09}
Wittenmyer R.~A.,  Endl M.,  Cochran W.~D.,  Levison H.~F.,   Henry G.~W.,
  2009, ApJS, 182, 97

\bibitem[\protect\citeauthoryear{Wright}{Wright}{2005}]{Wright05}
Wright J.~T.,  2005, PASP, 117, 657

\bibitem[\protect\citeauthoryear{Wright \& Howard}{Wright \&
  Howard}{2009}]{WrightHoward09}
Wright J.~T.,  Howard A.~W.,  2009, ApJS, 182, 205

\bibitem[\protect\citeauthoryear{Zechmeister \& K{\"u}rster}{Zechmeister \&
  K{\"u}rster}{2009}]{ZechKur09}
Zechmeister M.,  K{\"u}rster M.,  2009, A\&A, 496, 577

\makeatother
\end{thebibliography}

\appendix

\bsp

\label{lastpage}

\end{document}